\documentclass[11pt]{article}
\usepackage{latexsym,amsmath,amssymb,theorem}

\newcommand{\QCBZ}{\mathsf{QCB_0}}
\newcommand{\CBZ}{\mathsf{CB_0}}
\newcommand{\Pomega}{{P\hspace*{-1pt}\omega}}
\newcommand{\bfSig}{\mathbf{\Sigma}}
\newcommand{\bfPi}{\mathbf{\Pi}}
\newcommand{\dom}{\mathit{dom}}
\newcommand{\rng}{\mathit{rng}}
\newcommand{\graph}{\mathit{graph}}
\newcommand{\sfC}{\mathsf{C}}
\newcommand{\sfD}{\mathsf{D}}
\newcommand{\calN}{\mathcal{N}}
\newcommand{\Nk}[1]{\mathbb{N}\langle#1\rangle}
\newcommand{\INu}{\mathbb{N}_\infty}

\newtheorem{lemma}{Lemma}[section]
\newtheorem{theorem}[lemma]{Theorem}
\newtheorem{corollary}[lemma]{Corollary}
\newtheorem{proposition}[lemma]{Proposition}

\theorembodyfont{\normalfont}

\newtheorem{remark}[lemma]{Remark}

\newtheorem{definition}[lemma]{Definition}

\begin{document}

\title{Some Hierarchies of $\QCBZ$-Spaces}

\author{Matthias Schr\"oder
\\University of Bundeswehr-Munich
\\Neubieberg, Germany
\\ and
\\Victor  Selivanov\thanks{Victor Selivanov has been supported by a Marie
International Research Staff Exchange Scheme Fellowship within the
7th European Community Framework Programme. Both authors have been
supported by DFG-RFBR Grant 436 RUS 113/1002/01.}\\A.P. Ershov
Institute of Informatics
Systems  SB RAS\\
Novisibirsk, Russia}

\date{}

\maketitle

\begin{abstract} We define and study hierarchies of topological spaces
induced by the classical Borel and Luzin hierarchies of sets. Our
hierarchies are divided into two classes: hierarchies of countably
based spaces induced by their embeddings into $\Pomega$, and
hierarchies of spaces (not necessarily countably based) induced by
their admissible representations. We concentrate on the
non-collapse property of the hierarchies and on the relationships
between hierarchies in the two classes.

\end{abstract}

%
%

\section{Introduction}\label{in}

A basic notion of Computable Analysis (CA) \cite{wei00} is
the notion of an {\em admissible representation} of a
topological space $X$. This is a partial continuous surjection
$\delta$ from the Baire space $\calN$ onto $X$
satisfying a certain universality property
(see Subsection~\ref{admiss} for some more details).
Such a representation of $X$ induces a reasonable
computability theory on $X$, and the class of admissibly
represented spaces is wide enough to include most spaces of
interest for Analysis or Numerical Mathematics. As shown by the
first author \cite{sch03}, this class coincides with the
class of the so-called $\QCBZ$-spaces, i.e. $T_0$-spaces which are
quotients of countably based spaces, and it forms a cartesian
closed category (with the continuous functions as morphisms).
Thus, among $\QCBZ$-spaces one meets many important
function spaces including the Kleene-Kreisel continuous functionals
\cite{kl59,kr59} interesting for several branches
of logic and computability theory.

Along with the mentioned nice properties of $\QCBZ$-spaces, this
class seems to be too broad to admit a deep understanding.
Hence, it makes sense to search for natural
subclasses of this class which still include ``practically''
important spaces but are (hopefully) easier to study. Interesting
examples of such subclasses are obtained if we consider, for each
level $\Gamma$ of the classical Borel or Luzin (projective)
hierarchies of Descriptive Set Theory (DST) \cite{ke95}, the
class of spaces which have an admissible representation of the
complexity $\Gamma$ (below we make this precise). The study of the
resulting Borel and Luzin
hierarchies of $\QCBZ$-spaces is one of
the aims of this paper.

Along with the hierarchies of $\QCBZ$-spaces, we will consider
Borel and Luzin hierarchies of countably based $T_0$-spaces
($\CBZ$-spaces for short) which are induced by the well-known fact
that any $\CBZ$-space may be embedded
in the algebraic domain $\Pomega$
of all subsets of $\omega$.
The precise definition depends crucially on the
possibility to find the ``right'' extensions of the classical
hierarchies from the class of Polish spaces to the arbitrary
spaces (including $\Pomega$). Such extensions were introduced
by the second author in \cite{s04a} and were studied in
\cite{s05,s05a,s06,s08,by09,br}. In particular, it was shown that
many properties of these hierarchies in $\omega$-continuous
domains resemble the properties of classical hierarchies in Polish
spaces \cite{ke95}.

Hierarchies of spaces obtained in this way turn out to be closely
related to the corresponding hierarchies of $\QCBZ$-spaces.
Moreover, among the first levels of the Borel hierarchy of
$\CBZ$-spaces we meet some classes of spaces which attracted
attention of several researches in the field of quasi-metric
spaces, in particular the class of quasi-Polish spaces. The class
of quasi-Polish spaces identified and studied in \cite{br} is a
good solution to the problem from \cite{s08} of finding a natural
class of spaces that includes the Polish spaces and the
$\omega$-continuous domains and has a reasonable DST.

In this paper we establish some basic properties of the
above-mentioned hierarchies of spaces. In particular, we show that
the hierarchies of spaces do not collapse, and that any level of
any hierarchy is closed under retracts. As main technical tools to
prove these results we use suitable generalizations of some
classical facts (e.g. of the injectivity property of $\Pomega$ and
of Lavrentyev's Theorem on extending partial homeomorphisms in
Polish spaces). Some of those generalizations might be also
interesting in their own right. We also show that the class of all
spaces in our hierarchies forms in a sense the smallest cartesian
closed category of $\QCBZ$-spaces containing the discrete space
$\omega$ of natural numbers, and establish the close relationship
of the Luzin hierarchy of $\QCBZ$-spaces to the continuous
functionals of finite type. Hence the class of all spaces in the
Luzin hierarchy of $\QCBZ$-spaces seems to be a reasonable
subclass of $\QCBZ$-spaces that contains most of spaces of
interest for CA, including the Kleene-Kreisel continuous functionals.

After recalling some notions and known facts in the next section,
we establish the main technical facts in Section \ref{technic}. In
Sections \ref{cb} and \ref{qcb} we introduce and study the
mentioned hierarchies of $\CBZ$-spaces and of $\QCBZ$-spaces. In
Section \ref{relat} we establish close relationships between
hierarchies of $\CBZ$-spaces to those of $\QCBZ$-spaces (namely,
any level of the $\CBZ$-hierarchies coincides with
the class of $\CBZ$-spaces in the corresponding level of
the corresponding $\QCBZ$-hierarchy). In Section \ref{functional}
we relate the Luzin hierarchy of $\QCBZ$-spaces to the
Kleene-Kreisel continuous functionals. In Section~\ref{categ} we
discuss the cartesian closedness of the corresponding categories,
and we conclude in Section~\ref{con}.

%
%

\section{Notation and Preliminaries}\label{sec:prelim}

\subsection{Notation}\label{subnot}

We freely use the standard set-theoretic notation like
$\dom(f),\rng(f)$ and $\graph(f)$ for the domain, range and graph of
a function $f$, respectively, $X\times Y$ for the Cartesian
product, $X \sqcup Y$ for the disjoint union of sets $X$
and $Y$, $Y^X$ for the set of functions $f \colon X\to Y$
(but in the case when $X,Y$ are $\QCBZ$-spaces we use the same
notation to denote the space of continuous functions
from $X$ to $Y$), and $P(X)$ for the set of all subsets of $X$.
For $A\subseteq X$, $\overline{A}$ denotes the complement
$X\setminus A$ of $A$ in $X$. We identify the set of natural
numbers with the first infinite ordinal $\omega$. The first
uncountable ordinal is denoted by $\omega_1$.
The notation $f:X\to Y$
(resp.\ $f:\subseteq X\to Y$) means that $f$ is a total (resp.\ a
partial) function from a set $X$ to a set $Y$.

\subsection{Topological Spaces}\label{sub:topspaces}

We assume the reader to be familiar with the basic notions of
topology. The collection of all open subsets of a topological space $X$
(i.e.\ the topology of $X$) is denoted by $\tau_X$;
for the underlying set of $X$ we will write $X$ in abuse of notation.
We will usually abbreviate ``topological space'' to ``space''.
Remember that a space is \emph{zero-dimensional} if it has
a basis of clopen sets.

A space $Y$ is called a \emph{(continuous) retract} of a space $X$ if there
are continuous functions $s:Y\to X$ and $r:X\to Y$ such that
composition $rs$ coincides with the identity function $id_Y$ on
$Y$. Such a pair of functions $(s,r)$ is called a
{\em section-retraction} pair. Note that the section $s$ is a
homeomorphism between $Y$ and the subspace $s(Y)=\{x\in X\mid
sr(x)=x\}$ of $X$, and $s^{-1}=r|_{s(Y)}$.

Let $\omega$ be the space of non-negative integers with the
discrete topology. Of course, the spaces
$\omega\times\omega=\omega^2$, and $\omega\sqcup\omega$ are
homeomorphic to $\omega$, the first homeomorphism is realized by
the Cantor pairing function $\langle \cdot,\cdot\rangle$.

Let ${\mathcal N}=\omega^\omega$ be the set of all infinite
sequences of natural numbers (i.e., of all functions $\xi \colon
\omega \to \omega$). Let $\omega^*$ be the set of finite sequences
of elements of $\omega$, including the empty sequence. For
$\sigma\in\omega^*$ and $\xi\in{\mathcal N}$, we write
$\sigma\sqsubseteq \xi$ to denote that $\sigma$ is an initial
segment of the sequence $\xi$. By $\sigma\xi=\sigma\cdot\xi$ we
denote the concatenation of $\sigma$ and $\xi$, and by
$\sigma\cdot\calN$ the set of all extensions of $\sigma$ in
$\calN$. For $x\in\calN$, we can write
$x=x(0)x(1)\dotsc$ where $x(i)\in\omega$ for each $i<\omega$. For
$x\in\calN$ and $n<\omega$, let $x^{<n}=x(0)\dotsc x(n-1)$
denote the initial segment of $x$ of length $n$. Notations in the
style of regular expressions like $0^\omega$, $0^\ast 1$ or
$0^m1^n$ have the obvious standard meaning.

By endowing $\calN$ with the product of the discrete
topologies on $\omega$, we obtain the so-called \emph{Baire space}.
The product topology coincides with the topology
generated by the collection of sets of the form
$\sigma\cdot\calN$ for $\sigma\in\omega^*$. The Baire space
is of primary importance for DST and CA. The importance stems from
the fact that many countable objects are coded straightforwardly
by elements of $\calN$, and it has very specific topological
properties. In particular, it is a perfect zero-dimensional space
and the spaces ${\mathcal N}^2$, ${\mathcal N}^\omega$,
$\omega\times{\mathcal N}={\mathcal N}\sqcup{\mathcal
N}\sqcup\cdots$ (endowed with the product topology) are all
homeomorphic to ${\mathcal N}$. Let $(x,y)\mapsto\langle
x,y\rangle$ be a homeomorphism between ${\mathcal N}^2$ and
${\mathcal N}$. The Baire space $\calN$ has
the following universality property for zero-dimensional $\CBZ$-spaces:

\begin{proposition} \cite[Theorems 1.1 and 7.8]{ke95} \label{p:inj1}
A topological space $X$ embeds into ${\mathcal N}$
iff $X$ is a zero-dimensional $\CBZ$-space.
\end{proposition}

The subspace ${\mathcal C}=2^\omega$ of ${\mathcal N}$ formed by
the infinite binary strings (endowed with the relative topology
inherited from $\calN$) is known as the {\em Cantor space}.
Along with $\calN$ and $\mathcal{C}$, the space
$\Pomega$ is of principal
importance for this paper. This is the space of subsets of
the natural numbers with the Scott topology on $(P(\omega);\subseteq)$.
The basic open sets of this topology are of the form
$\{A\subseteq\omega\mid F\subseteq A\}$, where $F$ ranges over the
finite subsets of $\omega$.

The importance of $\Pomega$ for
this paper is explained by its following well-known properties.
First, $\Pomega$ is universal for $\CBZ$-spaces.

\begin{proposition}\label{inj2}
 A topological space $X$ embeds into $\Pomega$ iff $X$ is a $\CBZ$-space.
\end{proposition}

{\bf Proof.} Since $\Pomega$ is a $\CBZ$-space, any space $X$
homeomorphic to a subspace $\Pomega$ is a $\CBZ$-space as well.
Conversely, if $\{\beta_i \,|\, i \in \omega\}$ is a base for a
countably-based $T_0$-space $X$, then the function $e\colon X \to
\Pomega$ mapping $x$ to $\{ i \in \omega \,|\, x \in \beta_i \}$
is a homeomorphic embedding of $X$ into $\Pomega$. $\Box$

The second property shows that $\Pomega$ is an injective object
in the category of all topological spaces.

\begin{proposition}\cite[Proposition 3.5]{g03}\label{inject}
 Let $Y$ be topological space and $X$ be a subspace of $Y$. Then any
 continuous function $f:X\to \Pomega$ can be extended to a
 continuous function $g:Y\to \Pomega$.
\end{proposition}

Completely metrisable spaces satisfy the following extension theorem
by Kuratowski.

\begin{proposition}\cite[Theorem 3.8]{ke95}\label{p:ext:completely}
 Let $X$ be a metrisable space and $Y$ be a completely metrisable space.
 Then any continuous function $f\colon A\to Y$ defined on a subset $A$ of $X$
 can be extended to a continuous function $g:G\to Y$,
 where $G$ is a $G_\delta$-subset of $Y$ with
 $A \subseteq G$. 
\end{proposition}


\subsection{Families of Pointclasses}\label{fam}

Here we recall a useful technical notion of a family of
pointclasses introduced in \cite{s11}.

A {\em pointclass} on $ X $ is simply a collection $
\mathbf{\Gamma}(X) $ of subsets of $ X $. We need the
following ``parameterized'' version of the notion of pointclass. A
{\em family of pointclasses} is a family $
\mathbf{\Gamma}=\{\mathbf{\Gamma}(X)\} $ indexed by arbitrary
topological spaces $X$ such that each $ \mathbf{\Gamma}(X) $ is
a pointclass on $ X $ and $ \mathbf{\Gamma} $ is closed under
continuous preimages, i.e.\ $ f^{-1}(A)\in\mathbf{\Gamma}(X) $
for every $ A\in\mathbf{\Gamma}(Y) $ and every continuous
function $ f \colon X\to Y $. In particular, any pointclass
$\mathbf{\Gamma}(X)$ in such a family is downward closed under the
Wadge reducibility on $ X $. Recall that $A\subseteq X$ is {\em
Wadge reducible} to $B\subseteq Y$ if $A=f^{-1}(B)$ for some
continuous function $f$ on $X$.

Trivial examples of families of pointclasses are
$\mathcal{E},\mathcal{F}$, where $\mathcal{E}(X)=\{\emptyset\}$
and $\mathcal{F}(X)=\{X\}$ for any space $X$. A basic example of a
family of pointclasses is given by the family
$\mathcal{O}=\{\tau_X\}$ of the topologies of all the spaces $X$.

Finally, we define some operations on families of pointclasses
which are relevant to hierarchy theory. First, the usual
set-theoretic operations will be applied to the families of
pointclasses pointwise: for example, the union $\bigcup_i
\mathbf{\Gamma}_i$ of the families of pointclasses
$\mathbf{\Gamma}_0,\mathbf{\Gamma}_1,\ldots$ is defined by
$(\bigcup_i\mathbf{\Gamma}_i)(X)=\bigcup_i\mathbf{\Gamma}_i(X)$.

Second, a large class of such operations is induced by the
set-theoretic operations of L.V. Kantorovich and E.M. Livenson
(see e.g. \cite{s11} for the general definition). Among them are
the operations $\mathbf{\Gamma}\mapsto\mathbf{\Gamma}_\sigma$
where $\mathbf{\Gamma}(X)_\sigma$ is the set of all countable
unions of sets in $\mathbf{\Gamma}(X)$, the operation
$\mathbf{\Gamma}\mapsto\mathbf{\Gamma}_c$ where
$\mathbf{\Gamma}(X)_c$ is the set of all complements of sets in
$\mathbf{\Gamma}(X)$, the operation
$\mathbf{\Gamma}\mapsto\mathbf{\Gamma}_d$ where
$\mathbf{\Gamma}(X)_d$ is the set of all differences of sets in
$\mathbf{\Gamma}(X)$, and the operation $\Gamma\mapsto\Gamma_p$
defined by $\Gamma_p(X)=\{pr_X(A)\mid A\in\Gamma(\calN\times
X)\}$ where $pr_X(A)=\{x\in X\mid \exists p\in\calN((p,x)\in
A)\}$ is the projection of $A\subseteq\calN\times X$ along
the axis $\calN$.

The next subsection contains some important examples of families
of pointclasses from hierarchy theory.


\subsection{Classical Hierarchies on Topological Spaces}\label{hier}

Let us recall the definition of the Borel hierarchy on arbitrary
spaces introduced in \cite{s04a}.

\begin{definition}\label{defbh} \rm
For $ \alpha<\omega_1 $, define the family of pointclasses
$\bfSig^0_\alpha = \{ \bfSig^0_\alpha(X)\}$ by
induction on $\alpha$ as follows:
$\bfSig^0_0(X)=\{\emptyset\}$, $\bfSig^0_1(X) = \tau_X$
is the collection of the open sets of $X$, $\bfSig^0_2(X) =
((\bfSig^0_1(X))_d)_\sigma$ is the collection of all
countable unions of differences of open sets, and
$\bfSig^0_\alpha(X) =
(\bigcup_{\beta<\alpha}(\bfSig^0_\beta(X))_c)_\sigma$ (for
$\alpha>2$) is the class of countable unions of sets in $
\bigcup_{\beta<\alpha}(\bfSig^0_\beta(X))_c$.

The sequence $\{\bfSig^0_\alpha(X)\}_{ \alpha<\omega_1}$ is
called \emph{the Borel hierarchy} of $X$.
We also let $\bfPi^0_\beta(X)= (\bfSig^0_\beta(X))_c $
and $\mathbf{\Delta}^0_\alpha(X) = \bfSig^0_\alpha(X) \cap
\bfPi^0_\alpha (X)$.
The classes
$\bfSig^0_\alpha(X),\bfPi^0_\alpha(X),{\bf\Delta}^0_\alpha(X)$
are called the \emph{levels} of the Borel hierarchy of $X$.
\end{definition}

By the definition and remarks at the end of the previous
subsection, any of
$\bfSig^0_\alpha,\bfPi^0_\alpha,{\bf\Delta}^0_\alpha$ is a
family of pointclasses. It is straightforward to check by
induction on $\alpha,\beta$ that using Definition \ref{defbh}
one has the following result.

\begin{proposition}\label{propbh}
For every topological space $X$ and for all $\alpha <\beta < \omega_1$,
$\bfSig^0_\alpha(X)\cup \bfPi^0_\alpha(X) \subseteq
{\bf\Delta}^0_\beta(X)$.
\end{proposition}

{\bf Remark.} Definition {\ref{defbh}} applies to all spaces $X$,
and Proposition {\ref{propbh}} holds true in the full
generality. Note that Definition {\ref{defbh}} differs from the
classical definition for Polish spaces
(see e.g.\ \cite[Section~11.B]{ke95}) only for the level 2,
and that for the case of Polish
spaces our definition of the Borel hierarchy is equivalent to the
classical one. Notice that the classical definition cannot be
applied in general to non metrizable spaces (like e.g.\ the non
discrete $\omega$-algebraic domains) precisely because the
inclusion $\bfSig^0_1 \subseteq \bfSig^0_2$ may fail.

Let $\{\bfSig^1_n(X)\}_{1\leq n<\omega}$ be
Luzin's projective hierarchy in $X$ (cf. \cite{br}). Using the
corresponding operation on families of pointclasses from the
previous subsection we have
$\bfSig^1_1(X)=(\bfPi^0_2(X))_p$ and
$\bfSig^1_{n+1}(X)=(\bfPi^1_n(X))_p$ for any $n\geq1$. Let
also $\bfSig^1_0=\bfPi^1_0=\mathbf{\Delta}^1_1$.
The reason why the definition of the first level of
the Luzin hierarchy is distinct from the classical definition
${\mathbf\Sigma}^1_1(X)=(\bfPi^0_1(X))_p$ for Polish spaces is
that the inclusion
${\mathbf\Sigma}^0_1(X)\subseteq(\bfPi^0_1(X))_p$ may fail in
general.

Any level $\bfSig^1_n,\bfPi^1_n,\mathbf{\Delta}^1_n$ of the Luzin
hierarchy is a family of pointclasses, and we have the natural
inclusions among them similar to the inclusions for levels of the
Borel hierarchy.

Levels of the introduced hierarchies in an arbitrary space have
closure properties similar to those known for the classical
hierarchies in Polish spaces, in particular:

\begin{proposition}\label{clos}
Any non-zero $\bfSig$-level of the Borel hierarchy on an arbitrary
topological space is closed under finite intersections, countable
unions and binary products (for the case of binary product this
means that $A\in\mathbf{\Sigma}^0_\alpha(X)$ and
$B\in\mathbf{\Sigma}^0_\alpha(Y)$ imply $A\times
B\in\mathbf{\Sigma}^0_\alpha(X\times Y)$).
Any non-zero $\bfSig$-level of the Luzin hierarchy in an arbitrary
space is closed under countable unions, countable intersections,
binary products and the projection along $\calN$-axis (for the
case of projection this means that
$A\in\mathbf{\Sigma}^1_n(\mathcal{N}\times X)$ implies
$pr_X(A)\in\mathbf{\Sigma}^1_n(X)$).
\end{proposition}

We will often use the following straightforward result which also
extends the corresponding well-known facts for the classical
hierarchies.

\begin{proposition}\label{sub}
 Let $X$ be a subspace of a topological space $Y$, $A\subseteq X$,
 and let $\Gamma$ be a $\bfSig$- or a $\bfPi$-level
 of the introduced hierarchies.
 Then $A\in\Gamma(X)$ iff $A=X\cap B$ for some $B\in\Gamma(Y)$.
 If
 in addition $X \in \Gamma(Y)$, then $A\in\Gamma(X)$ iff $A\in\Gamma(Y)$.
\end{proposition}

We will often cite the following topological complexity
of the equality test $EQ_X:=\{(x,x)\mid x\in X\}$
on a topological space $X$.

\begin{proposition}\label{equal}
If $X$ is a $\CBZ$-space then $EQ_X\in\bfPi^0_2(X\times X)$.
If $X$ is a Hausdorff space, then
$EQ_X \in \bfPi^0_1(X\times X)$.
\end{proposition}

{\bf Proof.}
 The first statement has been shown in \cite{br}.
 The second (well-known) statement follows from the fact that
 in the Hausdorff case the complement of $EQ_X$
 is equal to
 $\displaystyle\bigcup\{ U \times V \,|\,
    U,V \text{ open and disjoint} \}$
 and therefore open.
$\Box$

Let $\Gamma$ be a family of pointclasses and $X,Y$ be spaces. A
function $f:X\to Y$ is called {\em $\Gamma$-measurable} if
$f^{-1}(A)\in\Gamma(X)$ for each open set $A\subseteq Y$. Note
that the continuous functions coincide with the
$\bfSig^0_1$-measurable functions.

We will also use the following basic property of $\Gamma$-measurable
functions.
Note that addition $+$ on ordinal numbers is associative,
but not commutative.

\begin{proposition}\cite[Lemma 1]{by09}\label{meas}
 Let $X,Y$ be $\CBZ$-spaces,
 let $\alpha,\beta<\omega_1$ be ordinals,
 let $f:X\to Y$ be a $\bfSig^0_{1+\alpha}$-measurable function,
 and let $A\in\bfSig^0_{1+\beta}(Y)$ and
 $B\in\bfPi^0_{1+\beta}(Y)$.
 Then
 $f^{-1}(A)\in\bfSig^0_{1+\alpha+\beta}(X)$ and
 $f^{-1}(B)\in\bfPi^0_{1+\alpha+\beta}(X)$.
\end{proposition}

Notice that the case $(\alpha,\beta)=(0,0)$ denotes the fact that
the preimage of an open (closed) set under a continuous function
is open (resp.\ closed).


\subsection{Polish and quasi-Polish spaces}\label{subsectionpolish}

Remember that a space $ X $ is {\em Polish} if it is countably
based and metrisable with a metric $d$ such that $(X,d)$ is a
complete metric space. Important examples of Polish spaces are the
Baire space, the Cantor space, the space of reals $ \mathbb{R} $
and its Cartesian powers $ \mathbb{R}^n $ ($ n < \omega $), the
closed unit interval $ [0,1] $, the Hilbert cube $ [0,1]^\omega $
and the Hilbert space $ \mathbb{R}^\omega $.

Below we will cite the following result known as Lavrentyev's Theorem:

\begin{proposition}\cite[Theorem 3.9]{ke95}\label{lavren}
 Let $X,Y$ be Polish spaces,
 $A\subseteq X$, $B\subseteq Y$, and let $f$ be a homeomorphism of
 $A$ onto $B$ (equipped with the subspace topologies induced from $X$ and $Y$).
 Then there exist $A^\ast\in\bfPi^0_2(X)$,
 $B^\ast\in\bfPi^0_2(Y)$ and a homeomorphism $f^\ast$ of
 $A^\ast$ onto $B^\ast$ such that $A\subseteq A^\ast$, $B\subseteq
 B^\ast$, and $f^\ast|_A=f$.
\end{proposition}

A natural variant of the class of Polish spaces has recently
emerged. Given a set $ X $, call a function $ d $ from $ X\times X
$ to the nonnegative reals \emph{quasi-metric} whenever $ x=y $
iff $ d(x,y)=d(y,x)=0 $, and $ d(x,y)\leq d(x,z)+d(z,y) $ (but we
don't require $ d $ to be symmetric). In particular, every metric
is a quasi-metric. Every quasi-metric on $ X $ canonically induces
a topology on $ X $ which is denoted by $ \tau_d $, where $ \tau_d
$ is the topology generated by the open balls $ B_d(x,
\varepsilon) = \{ y \in x \mid d(x,y) < \varepsilon \} $ for $ x
\in X $ and $ 0 < \varepsilon \in \mathbb{R}$. A space $ X $ is
called \emph{quasi-metrisable} if there is a quasi-metric on $ X $
which generates its topology. If $ d $ is a quasi-metric on $ X $,
let $ \hat{d} $ be the metric on $ X $ defined by $ \hat{d}(x,y)=
\max\{d(x,y),d(y,x)\} $. A sequence $ \{ x_n \}_{ n < \omega}$ in
$X$ is called {\em$ d $-Cauchy sequence} if for every $
\varepsilon > 0 $ there is $ n_0 \in \omega $ such that $
d(x_n,x_m) < \varepsilon $ for all $ n_0 \leq n \leq m $. We say
that the quasi-metric $ d $ on $ X $ is \emph{complete} if every $
d $-Cauchy sequence converges with respect to $ \hat{d} $ (notice
that this definition is coherent with the notion of completeness
for a metric $ d $, as in this case $ \hat{d} = d $).

A $T_0$-space $X$ is called \emph{quasi-Polish} if it is
countably based and there is a complete quasi-metric on $X$ which
generates its topology. In particular, every Polish space is
quasi-Polish, but by \cite[Corollary 45]{br} also every
$\omega$-continuous domain is quasi-Polish.
De Brecht \cite{br} shows that there is a reasonable DST for the
quasi-Polish spaces which extends the classical DST for Polish
spaces \cite{ke95} and the DST for $\omega$-continuous domains
\cite{s06,s08} in many directions.

An important example of a quasi-Polish space is the space
$\Pomega$ equipped with the following quasi-metric $d$: if
$A\subseteq B$ then $d(A,B)=0$, otherwise $d(A,B)=1/2^a$ where $a$
is the smallest number in $A\setminus B$. Note that Proposition
\ref{inj2} implies that any $\CBZ$-space is quasi-metrisable.

De Brecht proved the following characterization of quasi-Polish spaces
(cf.\ \cite[Corollary 24]{br}):

\begin{proposition}\label{proppi2}
 A topological space is quasi-Polish iff it is homeomorphic
 to a ${\mathbf\Pi}^0_2$-subset of $\Pomega$
 (endowed with the relative topology inherited from $\Pomega$).
\end{proposition}


\subsection{Admissible Representations and $\QCBZ$-spaces}
\label{admiss}

A \emph{representation} of a space $X$ is a
surjection of a subspace of the Baire space $\calN$ onto $X$.
Usually it is denoted as a partial function from $\calN$ to $X$.
The notion of admissible representation is basic
in Computable Analysis (CA).
A representation $\delta$ of $X$ is \emph{admissible}, if
it is continuous and
any continuous function $\nu:Z \to X$ from a zero-dimensional $\CBZ$-space
$Z$ to $X$ is continuously reducible to $\delta$,
i.e. $\nu=\delta g$ for some continuous function $g:Z \to \calN$.
A topological space is \emph{admissibly representable} if it has
an admissible representation.

The notion of admissibility was introduced in \cite{kw85}
for representations of countably based spaces
(in a different but equivalent formulation)
and was extensively studied by many authors.
In \cite{bh02} a close
relation of admissible representations of countably based spaces
to open continuous representations was established.
In \cite{sch02,sch03} the notion was extended to non-countably based spaces
and a nice characterization of the admissibly represented spaces
was achieved. Namely, the admissibly represented
sequential topological spaces
coincide with the $\QCBZ$-spaces, i.e., $T_0$-spaces which are
topological quotients of countably based spaces.

The category $\QCBZ$ of $\QCBZ$-spaces as objects and
continuous functions
as morphisms is known to be cartesian closed (cf.\ \cite{ELS04,sch03}).
The exponential $Y^X$ to $\QCBZ$-spaces $X,Y$ has
the set of continuous functions from $X$ to $Y$ as the underlying set,
and its topology is the sequentialization of the compact-open topology
on $Y^X$.
By the \emph{sequentialization} of a topology $\tau$ we mean
the family of all sequentially open sets pertaining to this topology.
(Remember that \emph{sequentially open} sets are defined to be the complements
of the sets that are closed under forming limits of converging sequences.)
The sequentialization of $\tau$ is finer than or equal to $\tau$.
The topology of the $\QCBZ$-product to $X$ and $Y$,
which we denote by $X \times Y$,
is the sequentialization of the classical Tychonoff topology
on the cartesian product of the underlying sets of $X$ and $Y$.
So products and exponentials in $\QCBZ$ are formed in the same way
as in its supercategory $\mathsf{Seq}$ of sequential topological spaces.

From \cite{br} it follows that admissible total representations
are closely related to quasi-Polish spaces. The following
assertion is contained among the results in \cite{br}.

\begin{proposition}\label{adm} For any $\CBZ$-space $X$
 the following statements are equivalent:
 \begin{description}
 \item[\rm(1)\,] $X$ is quasi-Polish.
 \item[\rm(2)\,] $X$ has an open continuous total representation.
 \item[\rm(3)\,] $X$ has an admissible total representation.
 \item[\rm(4)\,] $X$ has an admissible representation whose domain is a Polish space.
\end{description}
\end{proposition}

We will also cite the following facts from \cite{sch02,sch03}.

\begin{proposition}\label{real}
 Let $\delta$ and $\gamma$ be admissible representations of $\QCBZ$-spaces
 $X$ and $Y$, respectively.
 Then $f:X\to Y$ is continuous
 iff $f\delta=\gamma g$ for some partial continuous function
 $g$ on $\calN$.
\end{proposition}

Note that if $u:\subseteq\calN^2\to\calN$ is
continuous and $p\in\calN$ then $u_p=\lambda x.u(p,x)$ is a
partial continuous function on $\calN$.

\begin{proposition}\label{p:univ}
There is a partial continuous function
$u\colon\subseteq\calN^2\to\calN$ such that
$\dom(u)\in\bfPi^0_2(\calN^2)$ and for any partial
continuous function $g$ on $\calN$ there is
some $p\in\calN$ such that $u_p$ is an extension of $g$.
\end{proposition}

The function $u$ in Proposition~\ref{p:univ} can be chosen
as the application operator of the Second Kleene Algebra.
We use it to describe the construction of admissible representations
for function spaces formed in $\QCBZ$
(cf.\ \cite{sch03,wei00}).

\begin{proposition}\label{fspace}
 Let $\delta$ and $\gamma$ be admissible representations
 for $\QCBZ$-spaces $X$ and $Y$, respectively.
 Then admissible representations $[\delta\times\gamma]$
 for the $\QCBZ$-product $X \times Y$
 and $[\delta\to\gamma]$ for the $\QCBZ$-exponential $Y^X$
 can be defined by:
 \[
   \text{ $[\delta\times\gamma](\langle p,q \rangle)=(x,y)$
          iff $\delta(p)=x \wedge \gamma(q)=y$}
   \quad\mathrm{and}\quad
   \text{$[\delta\to\gamma](p)=f$ iff $f\delta=\gamma u_p$.}
 \]
 for $p,q \in \calN$, $x \in X$, $y \in Y$,
 and $f\colon X \to Y$.
\end{proposition}

De Brecht and Yamamoto showed the following property
of subsets of admissibly represented countably-based spaces.

\begin{proposition}\label{p:BrYa} \cite[Corollary 3]{by09}.
 Let $\delta$ be an admissible representation
 of a countably-based $T_0$-space $X$,
 let $A \subseteq X$ and let $1 \leq \alpha < \omega_1$.
 Then $A \in \bfSig^0_\alpha(X)$
 iff $\delta^{-1}(A) \in \bfSig^0_\alpha(\dom(\delta))$.
\end{proposition}

%
%

\section{Main Technical Facts}\label{technic}

In this section we prove a couple of facts that serve as main
technical tools in the sequel, but some of these facts might be
also of independent interest.

The first result generalizes the well-known fact that $\Pomega$ is
an injective space (see Proposition \ref{inj2}), because the
continuous functions coincide with the
$\bfSig^0_1$-measurable functions.

\begin{theorem}\label{inj}
 Let $Y$ be a topological space, let $X$ be a subspace of $Y$,
 and let
 $\Gamma\in\{\bfSig^0_\alpha,
 \bfSig^1_n,\bfPi^1_n\mid
 1\leq\alpha<\omega_1,1\leq n<\omega\}$.
 Then any $\Gamma$-measurable function $f:X\to \Pomega$
 can be extended to
 a $\Gamma$-measurable function $g:Y\to \Pomega$.
\end{theorem}

{\bf Proof.} For any $n<\omega$, the set
${\uparrow}\{n\}=\{A\subseteq\omega\mid n\in A\}$ is open in
$\Pomega$, hence $f^{-1}({\uparrow}\{n\})\in\Gamma(X)$, hence
$f^{-1}({\uparrow}\{n\})=X\cap A_n$ for some $A_n\in\Gamma(Y)$ by
Proposition \ref{sub}. Define the function $g:Y\to \Pomega$ by
$g(y)=\{n\mid y\in A_n\}$. Then for any $y\in Y$ we have
\[ y\in A_n
  \Leftrightarrow n\in g(y)
  \Leftrightarrow g(y)\in{\uparrow}\{n\}
  \Leftrightarrow y\in g^{-1}({\uparrow}\{n\}),
\]
hence $g^{-1}({\uparrow}\{n\})=A_n\in\Gamma(Y)$.
Since $\{{\uparrow}\{n\}\mid n\in\omega\}$ is a
subbasis in $\Pomega$ and $\Gamma(Y)$ is closed under finite
intersection and countable union by Proposition \ref{clos}, $g$ is
$\Gamma$-measurable.
If $y\in X$ then we have
\[ n\in f(y)
  \Leftrightarrow f(y)\in{\uparrow}\{n\}
  \Leftrightarrow y\in f^{-1}({\uparrow}\{n\})
  \Leftrightarrow y\in A_n
  \Leftrightarrow n\in g(y),
\]
 hence $g(y)=f(y)$.
$\Box$

The second result is a remote relative of Lavrentyev's Theorem (cf.
Proposition \ref{lavren}).

\begin{theorem}\label{lavrPo}
 Let $A,B\subseteq \Pomega$ be subspaces of $\Pomega$,
 $\alpha,\beta<\omega_1$, and let $f:A\to B$ be a
 $\bfSig^0_{1+\alpha}$-measurable bijection such that its
 inverse $g=f^{-1}$ is $\bfSig^0_{1+\beta}$-measurable.
 Then there exist
 $A^\ast\in\bfPi^0_{1+\alpha+\mu+1}(\Pomega)$,
 $B^\ast\in\bfPi^0_{1+\beta +\mu+1}(\Pomega)$ (where
 $\mu=max\{\alpha,\beta\}$) and a
 $\bfSig^0_{1+\alpha}$-measurable bijection
 $f^\ast:A^\ast\to B^\ast$ such that $g^\ast=f^{\ast-1}$ is
 $\bfSig^0_{1+\beta}$-measurable, $A\subseteq A^\ast$,
 $B\subseteq B^\ast$, $f^\ast|_A=f$ and $g^\ast|_B=g$.
\end{theorem}

{\bf Proof.} By Theorem \ref{inj}, there are a
$\bfSig^0_{1+\alpha}$-measurable extension $f_1:\Pomega\to
\Pomega$ of $f$ and a $\bfSig^0_{1+\beta}$-measurable extension
$g_1:\Pomega\to \Pomega$ of $g$. Consider the set
\[ S=\{(x,y)\in \Pomega\times \Pomega\mid f_1(x)=y\wedge
   x=g_1(y)\}.
\]
Clearly, $S=S_1\cap S_2$ where $S_1=\{(x,y)\mid
f_1(x)=y\}$ and $S_2=\{(x,y)\mid x=g_1(y)\}$. Since
$EQ_{\Pomega}\in\bfPi^0_2(\Pomega\times \Pomega)$ by
Proposition \ref{equal} and $S_1$ is the preimage of
$EQ_{\Pomega}$ under the $\bfSig^0_{1+\alpha}$-measurable
function $(x,y)\mapsto(f_1(x),y)$,
$S_1\in\bfPi^0_{1+\alpha+1}(\Pomega\times \Pomega)$ by
Proposition \ref{meas}. Similarly,
$S_2\in\bfPi^0_{1+\beta+1}(\Pomega\times \Pomega)$ and
therefore $S\in\bfPi^0_{1+\mu+1} (\Pomega\times \Pomega)$
by Propositions \ref{propbh} and \ref{clos}.

Now let $A^\ast=\{x\in \Pomega\mid (x,f_1(x))\in S\}$ and
$B^\ast=\{y\in \Pomega\mid (g_1(y),y)\in S\}$. Since the function
$x\mapsto(x,f_1(x))$ is $\bfSig^0_{1+\alpha}$-measurable
and $A^\ast$ is the preimage of $S$ under this function,
$A^\ast\in\bfPi^0_{1+\alpha+\mu+1}(\Pomega)$ by Proposition
\ref{meas}. Similarly,
$B^\ast\in\bfPi^0_{1+\beta+\mu+1}(\Pomega)$. Then the sets
$A^\ast,B^\ast$ and the functions $f^\ast=f_1|_{A^\ast}$,
$g^\ast=g_1|_{B^\ast}$ have the desired properties.
$\Box$

The next fact is a special case of the previous theorem.

\begin{corollary}\label{lavrPo1}
 If we take in the previous theorem $\alpha=0$ and $\beta=1$
 then we obtain $A^\ast\in\bfPi^0_3(\Pomega)$ and
 $B^\ast\in\bfPi^0_4(\Pomega)$.
\end{corollary}

Let us introduce one of the main notions of this paper.

\begin{definition}\label{def:GammaSpace} \rm
Let $\Gamma$ be a family of pointclasses. A topological space $X$
is called a \emph{$\Gamma$-space} if $X$ is homeomorphic to a subspace
$A\subseteq \Pomega$ with $A\in\Gamma(\Pomega)$. The class of all
$\Gamma$-spaces is denoted $\CBZ(\Gamma)$.
\end{definition}

The third result of this section extends Corollary 24 in \cite{br}
(which is obtained when $\Gamma=\bfPi^0_2$).

\begin{proposition}\label{p:ext}
Let $\Gamma\in\{\bfPi^0_2,\bfSig^0_\alpha, \bfPi^0_\alpha,
\bfSig^1_n, \bfPi^1_n\mid
3\leq\alpha<\omega_1,1\leq n<\omega\}$ and $f:\subseteq X\to Y$ be
a partial continuous function from a topological space $X$
to a $\Gamma$-space $Y$.
Then there is a continuous extension $g:\subseteq X\to Y$ of
$f$ with $\dom(g)\in\Gamma(X)$.
\end{proposition}

{\bf Proof.} Without loss of generality we assume
$Y\in\Gamma(\Pomega)$, so in particular $f:\subseteq X\to
\Pomega$. By Proposition \ref{inject}, there is a total continuous
extension $h: X\to \Pomega$ of $f$. Since $Y\in\Gamma(\Pomega)$,
$h^{-1}(Y)\in\Gamma(X)$. Since $\dom(f)\subseteq h^{-1}(Y)$, we
can take the restriction of $h$ to $h^{-1}(Y)$ as the desired
function $g$. $\Box$

The fourth result of this section is the following extension of
Lavrentyev's Theorem (see Proposition \ref{lavren}). For
$\Gamma=\bfPi^0_2$ the result gives the extension of
Lavrentyev's Theorem to quasi-Polish spaces (cf. \cite{br}), and
Lavrentyev's Theorem is obtained if we restrict the last
fact to Polish spaces.

\begin{theorem}\label{th:generalized:lav}
Let $\Gamma\in\{\bfPi^0_2,\bfSig^0_\alpha, \bfPi^0_\alpha,
\bfSig^1_n, \bfPi^1_n\mid
3\leq\alpha<\omega_1,1\leq n<\omega\}$, $X,Y$ be $\Gamma$-spaces,
$A\subseteq X$, $B\subseteq Y$, and let $f$ be a homeomorphism of
$A$ onto $B$. Then there exist $A^\ast\in\Gamma(X)$,
$B^\ast\in\Gamma(Y)$ and a homeomorphism $f^\ast$ of $A^\ast$
onto $B^\ast$ such that $A\subseteq A^\ast$, $B\subseteq B^\ast$,
and $f^\ast|_A=f$.
\end{theorem}

{\bf Proof.} Let $g=f^{-1}$. By Proposition \ref{p:ext} there
exist $A_1\in\Gamma(X)$, $B_1\in\Gamma(Y)$ and continuous
functions $f_1:A_1\to Y$, $g_1:B_1\to X$ such that $A\subseteq
A_1$, $B\subseteq B_1$, $f_1|_A=f$ and $g_1|_B=g$. Let
\[
 S=\{(x,y)\in
  A_1\times B_1\mid f_1(x)=y\wedge x=g_1(y)\}.
\]
Then $\graph(f)\subseteq S$ and $S\in\Gamma(X\times Y)$,
because $A_1\times B_1\in\Gamma(X\times Y)$,
$EQ_X\in\bfPi^0_2(X\times X)$ and
$EQ_Y\in\bfPi^0_2(Y\times Y)$ by Proposition \ref{equal},
$(x,y)\mapsto(f_1(x),y))$ is a continuous function from $A_1\times
B_1$ to $Y\times Y$ and $(x,y)\mapsto(x,g_1(y))$ is a continuous
function from $A_1\times B_1$ to $X\times X$.

Let now
\[
 A^\ast=\{x\in A_1\mid \exists y((x,y)\in A_1\times B_1)\}
 ,\;
 B^\ast=\{y\in B_1\mid \exists x((x,y)\in A_1\times B_1)\},
\]
$f^\ast=f_1|_A$ and $g^\ast=g_1|_B$. Then $A\subseteq
A^\ast$, $B\subseteq B^\ast$, $f^\ast|_A=f$, $g^\ast|_B=g$ and
$g^\ast=f^{\ast-1}$ (hence $f^\ast$ is a homeomorphism of $A^\ast$
onto $B^\ast$). Since
\[
 A^\ast=\{x\in A_1\mid (x,f_1(x))\in A_1\times B_1\}
 \text{ and }
 B^\ast=\{y\in B_1\mid (g_1(y),y)\in A_1\times B_1\},
\]
$A^\ast\in\Gamma(A_1)$ and $B^\ast\in\Gamma(B_1)$.
By Proposition \ref{sub} we have
$A^\ast\in\Gamma(X)$ and $B^\ast\in\Gamma(Y)$.
$\Box$

Finally, we give a natural version of the previous theorem related
to retracts, although this version is not used in the sequel.

\begin{proposition}\label{lavret}
 Let $\Gamma,X,Y,A,B$ be as in the previous theorem and let
$r:A\to B$ and $s:B\to A$ be continuous functions such that
$rs=id_B$ (hence, $B$ is a retract of $A$). Then there exist
$A^\ast\in\Gamma(X)$, $B^\ast\in\Gamma(Y)$ and continuous
functions $r^\ast:A^\ast\to B^\ast$, $s^\ast:B^\ast\to A^\ast$
such that $A\subseteq A^\ast$, $B\subseteq B^\ast$, $r^\ast|_A=r$,
$s^\ast|_B=s$ and $r^\ast s^\ast=id_{B^\ast}$ (hence, $B^\ast$ is
a retract of $A^\ast$).
\end{proposition}

{\bf Proof.} By Proposition \ref{p:ext} there exist
$A_1\in\Gamma(X)$, $B_1\in\Gamma(Y)$ and continuous functions
$r_1:A_1\to Y$, $s_1:B_1\to X$ such that $A\subseteq A_1$,
$B\subseteq B_1$, $r_1|_A=r$ and $s_1|_B=s$. Let
\[
 S=\{(x,y)\in A_1\times B_1\mid r_1(x)=y=r_1s_1(y)\}
\]
and
\[
 A^\ast=\{x\in A_1\mid (x,r_1(x)) \in S\},\;
 B^\ast=\{y\in B_1\mid (s_1(y),y) \in S\},
\]
$r^\ast=r_1|_{A^\ast}$ and $s^\ast=s_1|_{B^\ast}$.
Similarly to the previous proof one checks
that these objects have the desired properties.
$\Box$

%
%

\section{Hierarchies of $\CBZ$-Spaces}\label{cb}

Here we introduce and study natural hierarchies of $\CBZ$-spaces
(see Definition~\ref{def:GammaSpace})
as well as the Borel hierarchy and the Luzin hierarchy
on $\Pomega$ discussed in Subsection \ref{hier}.

\pagebreak[3]
\begin{definition}\label{hiercb}
\begin{description}
\item[\rm(1)\,]
 By the \emph{Borel hierarchy} of $\CBZ$-spaces we mean the sequence of classes
 $\{\CBZ(\bfSig^0_\alpha)\}_{\alpha<\omega_1}$.
 By \emph{levels} of this hierarchy we mean the classes
 $\CBZ(\bfSig^0_\alpha)$ as well as the classes
 $\CBZ(\bfPi^0_\alpha)$ and
 $\CBZ(\mathbf{\Delta}^0_\alpha)$.
\item[\rm(2)\,]
 By the \emph{Luzin hierarchy} of $\CBZ$-spaces we mean the sequence of classes
 $\{\CBZ(\bfSig^1_n)\}_{n<\omega}$. By \emph{levels} of this
 hierarchy we mean the classes $\CBZ(\bfSig^1_n)$ as well
 as the classes $\CBZ(\bfPi^1_n)$ and
 $\CBZ(\mathbf{\Delta}^1_n)$.
\end{description}
\end{definition}

Note that the Borel hierarchy of spaces includes some natural classes
of spaces identified earlier. E.g., by Proposition \ref{proppi2}
$\CBZ(\bfPi^0_2)$ coincides with the class of quasi-Polish
spaces and, by Corollary 33 in \cite{br}, $\CBZ(\bfPi^0_3)$
coincides with the class of $\CBZ$-spaces that admit a compatible
bicomplete quasi-metric in the sense of \cite{jk98}.

Obviously, for any families of pointclasses $\Gamma,\Delta$ we
have: if $\Gamma\subseteq\Delta$ (i.e.
$\Gamma(X)\subseteq\Delta(X)$ for each space $X$) then
$\CBZ(\Gamma)\subseteq \CBZ(\Delta)$. Therefore, we have the
natural inclusions for levels of the introduced hierarchies, in
particular $\CBZ(\mathbf{\Delta}^0_\alpha)\subseteq
\CBZ(\bfSig^0_\alpha)\cap \CBZ(\bfPi^0_\alpha)$ and
$\CBZ(\bfSig^0_\alpha)\cup \CBZ(\bfPi^0_\alpha)\subseteq
\CBZ(\mathbf{\Delta}^0_\beta)$ for all $\alpha<\beta<\omega_1$.
Below we establish some basic properties of the Borel and the
Luzin hierarchy of spaces, but first we show that most of levels
of the introduced hierarchies of spaces are closed under retracts
(for $\bfPi^0_2$ the result is already known from \cite{br}).

\begin{proposition}\label{retr}
 Let $\Gamma\in\{\bfPi^0_2,\bfSig^0_\alpha, \bfPi^0_\alpha,
 \bfSig^1_n, \bfPi^1_n\mid 3\leq\alpha<\omega_1,1\leq n<\omega\}$.
 Then any retract of a $\Gamma$-space is a $\Gamma$-space.
\end{proposition}

{\bf Proof.} Let $Y$ be a retract of a $\Gamma$-space $X$ via a
section-retraction pair $(s,r)$ of continuous functions; we have
to show that $Y$ is a $\Gamma$-space. We may assume without loss
of generality that $X\in\Gamma(\Pomega)$. Since $s(Y)=\{x\in X\mid
sr(x)=x\}$, $s(Y)$ is the preimage of $EQ_X$ under the continuous
function $x\mapsto(sr(x),x)$. Since $EQ_X\in\bfPi^0_2(X\times X)$
by Proposition \ref{equal}, $s(Y)\in\bfPi^0_2(X)$ and thus
$s(Y)\in\Gamma(X)$ by Proposition \ref{propbh}. Proposition
\ref{sub} yields $s(Y)\in\Gamma(\Pomega)$. Since $Y$ is
homeomorphic to $s(Y)$, $Y$ is a $\Gamma$-space. $\Box$

Now we establish an interesting property of the introduced
hierarchies of spaces which implies the non-collapse property.

\begin{proposition}\label{p:delta}
 For any countable ordinal $\alpha\geq 2$,
 $\CBZ(\bfSig^0_\alpha)\cap
 \CBZ(\bfPi^0_\alpha)=\CBZ(\mathbf{\Delta}^0_\alpha)$. For
 any positive integer $n$, $\CBZ(\bfSig^1_n)\cap
 \CBZ(\bfPi^1_n)=\CBZ(\mathbf{\Delta}^1_n)$.
\end{proposition}

This proposition is based on the following lemma.

\begin{lemma}\label{l:CB(Gamma)=Gamma(Pomega)}
 Let $\Gamma\in
   \{\bfPi^0_2,\bfSig^0_\alpha, \bfPi^0_\alpha, \bfSig^1_n, \bfPi^1_n
    \mid 3\leq\alpha<\omega_1,1\leq n<\omega\}$
 and let $X$ be a subspace of $\Pomega$.
 Then $X \in \CBZ(\Gamma)$ iff $X \in \Gamma(\Pomega)$.
\end{lemma}

{\bf Proof.}
 If the underlying set of $X$ is in $\Gamma(\Pomega)$,
 then $X$ is a $\Gamma$-space by Definition~\ref{def:GammaSpace}.
 Conversely, let $X$ be a $\Gamma$-space.
 Then there is some $B \in \Gamma(\Pomega)$ such that
 $B$ (endowed with the subspace topology inherited from $\Pomega$)
 is homeomorphic to $X$.
 Let $h\colon X \to B$ be a homeomorphism.
 We apply Theorem~\ref{th:generalized:lav} to extend $h$
 to a homeomorphism $h^*\colon X^* \to B^*$,
 where $X^*$ and $B^*$ are sets in $\Gamma(\Pomega)$
 with $X \subseteq X^*$ and $B \subseteq B^*$.
 By Proposition~\ref{sub} we have $B \in \Gamma(B^*)$.
 Hence $X \in \Gamma(X^*)$, because $X =(h^*)^{-1}(B)$.
 Proposition~\ref{sub} yields $X \in \Gamma(\Pomega)$.
$\Box$

Now we are ready to give the proof of Proposition~\ref{p:delta}.

{\bf Proof.} The inclusions from right to left are obvious. It
remains to check $\CBZ(\Gamma)\cap \CBZ(\Gamma_c)\subseteq
\CBZ(\Gamma\cap\Gamma_c)$ for each
$\Gamma\in\{\bfPi^0_\alpha, \bfPi^1_n \mid
 2\leq\alpha<\omega_1,1\leq n<\omega\}$.
Let $Z\in \CBZ(\Gamma)\cap \CBZ(\Gamma_c)$. Then $Z$ is
homeomorphic to some subspaces $A,B$ of $\Pomega$ with
$A\in\Gamma(\Pomega)$ and $B\in\Gamma_c(\Pomega)$. Hence $B$ is a
$\Gamma$-space by being homeomorphic to $A$.
Lemma~\ref{l:CB(Gamma)=Gamma(Pomega)} yields $B \in
\Gamma(\Pomega)$. Therefore $B\in(\Gamma\cap\Gamma_c)(\Pomega)$
and hence $Z\in \CBZ(\Gamma\cap\Gamma_c)$.
$\Box$

\begin{corollary}\label{ncol}
 The Borel hierarchy and the Luzin hierarchy of $\CBZ$-spaces do not collapse.
 More precisely $\CBZ(\bfSig^0_\alpha)\not\subseteq
 \CBZ(\bfPi^0_\alpha)$ for each countable ordinal
 $\alpha\geq2$,
 and $\CBZ(\bfSig^1_n)\not\subseteq \CBZ(\bfPi^1_n)$
 for each positive integer~$n$.
\end{corollary}

{\bf Proof.} Proofs for both hierarchies are similar, so consider
only the Borel hierarchy. According to \cite{s05a}, there is a set
$A$ in $\bfSig^0_\alpha(\Pomega)\setminus\bfPi^0_\alpha(\Pomega)$.
The space $A$ (with the topology induced from $\Pomega$) is
obviously a $\bfSig^0_\alpha$-space. If it were a
$\bfPi^0_\alpha$-space, then by
Lemma~\ref{l:CB(Gamma)=Gamma(Pomega)} we would have
$A\in\bfPi^0_\alpha(\Pomega)$, a contradiction. $\Box$


\section{Hierarchies of $\QCBZ$-Spaces}\label{qcb}

As shown in \cite{sch03}, any $\QCBZ$-space has an admissible representation.
Here we introduce and study natural
hierarchies of $\QCBZ$-spaces induced by this fact.
For any representation $\delta$ of a space $X$, let
$\mathit{EQ}(\delta):=\{(p,q) \in \calN^2 \mid p,q\in
\dom(\delta)\wedge\delta(p)=\delta(q)\}$.

\begin{definition}\label{hierqcb}
\begin{description}
 \item[\rm(1)\,]
  Let $\Gamma$ be a family of pointclasses. A topological
  space $X$ is called \emph{$\Gamma$-representable} if $X$ has an
  admissible representation $\delta$ with
  $\mathit{EQ}(\delta)\in\Gamma(\calN\times\calN)$. The class of
  all $\Gamma$-representable spaces is denoted $\QCBZ(\Gamma)$.
 \item[\rm(2)\,]
  By the \emph{Borel hierarchy} of $\QCBZ$-spaces we mean the sequence
  $\{\QCBZ(\bfSig^0_\alpha)\}_{\alpha<\omega_1}$.
  By \emph{levels} of this hierarchy we mean the classes
  $\QCBZ(\bfSig^0_\alpha)$ as well as the classes
  $\QCBZ(\bfPi^0_\alpha)$ and
  $\QCBZ(\mathbf{\Delta}^0_\alpha)$.
 \item[\rm(3)\,]
  By the \emph{Luzin hierarchy} of $\QCBZ$-spaces we mean the sequence
  $\{\QCBZ(\bfSig^1_n)\}_{n<\omega}$.
  By \emph{levels} of this
  hierarchy we mean the classes $\QCBZ(\bfSig^1_n)$ as well
  as the classes $\QCBZ(\bfPi^1_n)$ and
  $\QCBZ(\mathbf{\Delta}^1_n)$.
\end{description}
\end{definition}

The next assertion establishes an equivalent simpler definition of
most levels for the case of $\CBZ$-spaces.

\begin{proposition}\label{p:equiv:Gamma-representable}
\begin{description}
\item[\rm(1)\,]
 Let
 $\Gamma\in\{\bfPi^0_2,\bfSig^0_\alpha, \bfPi^0_\alpha,\bfSig^1_n, \bfPi^1_n
 \mid 3\leq\alpha<\omega_1,1\leq n<\omega\}$ and let $X$ be a
 $\CBZ$-space. Then $X$ is $\Gamma$-representable iff $X$ has an
 admissible representation $\delta$ with
 $\dom(\delta)\in\Gamma(\calN)$.
\item[\rm(2)\,]
 Let
 $\Gamma\in\{\bfPi^0_1,\bfSig^0_\alpha, \bfPi^0_\alpha,\bfSig^1_n, \bfPi^1_n
 \mid 2\leq\alpha<\omega_1,1\leq n<\omega\}$
 and let $X$ be a Hausdorff space.
 Then $X$ is $\Gamma$-representable
 iff $X$ has an admissible representation $\delta$ with
 $\dom(\delta)\in \Gamma(\calN)$.
\end{description}
\end{proposition}

{\bf Proof.} The only-if-part of the first statement holds for any
topological space $X$ and any family of pointclasses $\Gamma$.
Indeed, let $\delta$ be an admissible representation of $X$ with
$\mathit{EQ}(\delta)\in\Gamma(\calN\times\calN)$. Since
$\dom(\delta)$ is the preimage of $\mathit{EQ}(\delta)$ under the
continuous function $x\mapsto(x,x)$,
$\dom(\delta)\in\Gamma(\calN)$.

Conversely, let $\delta$ be an admissible representation of $X$
with $\dom(\delta)\in\Gamma(\calN)$.
By Proposition~\ref{equal},
$EQ_X \in \bfPi^0_2(X \times X)$,
so $\mathit{EQ}(\delta)$ is a $\bfPi^0_2$-set
in $\dom(\delta)\times\dom(\delta)$ by the continuity of $\delta$.
The set $\dom(\delta)\times\dom(\delta)$ is a $\Gamma$-subset
of $\calN\times\calN$
by being the intersection of the sets
$\{(x,y)\in\calN\times\calN\mid x\in
\dom(\delta)\}\in\Gamma(\calN\times\calN)$
and
$\{(x,y)\in\calN\times\calN\mid y\in
\dom(\delta)\}\in\Gamma(\calN\times\calN)$.
By Propositions~\ref{propbh} and~\ref{sub}
we obtain $\mathit{EQ}(\delta)\in\Gamma(\calN\times\calN)$.

The second statement follows similarly by taking into account
that $EQ_X$ is closed in $X \times X$ by Proposition~\ref{equal},
if $X$ is a Hausdorff space.
$\Box$

The main advantage of the hierarchies of this section
over the hierarchies from the previous section is that they
include many natural non-countably based spaces, in particular the
spaces of Kleene-Kreisel continuous functionals,
as we will see later.

Obviously, for any families of pointclasses $\Gamma,\Delta$ we
have: $\Gamma\subseteq\Delta$ implies $\QCBZ(\Gamma)\subseteq
\QCBZ(\Delta)$, hence we have the natural inclusions for levels of
the introduced hierarchies, in particular
$\QCBZ(\mathbf{\Delta}^0_\alpha)\subseteq
\QCBZ(\bfSig^0_\alpha)\cap \QCBZ(\bfPi^0_\alpha)$
and $\QCBZ(\bfSig^0_\alpha)\cup
\QCBZ(\bfPi^0_\alpha)\subseteq
\QCBZ(\mathbf{\Delta}^0_\beta)$ for all $\alpha<\beta<\omega_1$.
We will show that these hierarchies do not collapse, but first we
establish the closure of the levels under retracts.

\begin{proposition}\label{retr1}
 Let
 $\Gamma\in\{\bfSig^0_\alpha, \bfPi^0_\alpha,
 \bfSig^1_n, \bfPi^1_n\mid
 1\leq\alpha<\omega_1,1\leq n<\omega\}$.
 Then any retract of a
 $\Gamma$-representable space is a $\Gamma$-representable space.
\end{proposition}

{\bf Proof.} Let $Y$ be a retract of a $\Gamma$-representable
space $X$ via a section-retraction pair $(s,r)$ of continuous
functions; we have to show that $Y$ is a $\Gamma$-representable
space. Let $\delta$ be an admissible representation for $X$ with
$\mathit{EQ}(\delta)\in\Gamma(\calN\times\calN)$. Then $r\delta$
is clearly an admissible representation for $Y$, so it suffices to
show that $\mathit{EQ}(r\delta)\in\Gamma(\calN\times\calN)$. Since
$s$ is an injection, $\mathit{EQ}(r\delta)$ is the preimage of
$\mathit{EQ}(\delta)$ under the continuous function
$(x,y)\mapsto(sr(x),sr(y))$. Therefore,
$\mathit{EQ}(r\delta)\in\Gamma(\calN\times\calN)$. $\Box$

The following observation is well-known.

\begin{lemma}\label{l:retr2}
 Let $r\colon D \to Y$ be a continuous function from a subspace $D$
 of $\calN$ to a zero-dimensional $\CBZ$-space $Y$.
 Then $r$ (viewed as a partial function from $\calN$ to $Y$)
 is an admissible representation for $Y$
 iff there is a continuous function $s\colon Y \to D$
 satisfying $rs=\mathit{id}_Y$.
\end{lemma}

{\bf Proof.} Let $Y$ be a retract of $D$ via a section-retraction
pair $(s,r)$ of continuous functions. Then $r:\subseteq \calN \to
Y$ is admissible, because any continuous function $\nu:Z \to Y$
defined on a zero-dimensional $\CBZ$-space $Z$ is reducible to $r$
via $s\nu$.

Conversely, assume that $r \subseteq\colon \calN \to Y$
is an admissible representation of $Y$.
Then the identity function $id_Y$ is reducible to $r$
via some continuous function $s:Y\to D$.
Then $id_Y=r s$, hence $Y$ is a retract of $D$ via
the continuous section $s$ and the continuous retraction $r$.
$\Box$

Finally, we establish the non-collapse property of the introduced
hierarchies of spaces.

\begin{theorem}\label{ncol1}
 The Borel hierarchy and the Luzin hierarchy of $\QCBZ$-spaces
 do not collapse.
 More precisely, $\QCBZ(\bfSig^0_\alpha)\not\subseteq
 \QCBZ(\bfPi^0_\alpha)$ for each countable ordinal
 $\alpha\geq 2$,
 and
 $\QCBZ(\bfSig^1_n)\not\subseteq \QCBZ(\bfPi^1_n)$
  for each positive integer $n$.
\end{theorem}

{\bf Proof.} Proofs for both hierarchies are similar, so consider
only the Borel hierarchy. As is well-known \cite{ke95}, there is a
set $Y$ in $\bfSig^0_\alpha(\calN)\setminus
\bfPi^0_\alpha(\calN)$. Consider $Y$ as a subspace of $\calN$.
Since $id_Y$ is an admissible representation for $Y$, $Y\in
\QCBZ(\bfSig^0_\alpha)$. Suppose for a contradiction that $Y\in
\QCBZ(\bfPi^0_\alpha)$, so there is an admissible representation
$\delta:D\to Y$ for $Y$ with $D\in\bfPi^0_\alpha(\calN)$. By
Lemma~\ref{l:retr2} there is a continuous function $s:Y\to D$
satisfying $\delta s=id_Y$. Since $D$ is Hausdorff, $S=s(Y)$ is a
closed subset of $D$, hence $S \in \bfPi^0_\alpha(D)$ by
Proposition \ref{clos}. By Proposition~\ref{sub} we obtain
$S\in\bfPi^0_\alpha(\calN)$.

Since $\calN$ is a $\bfPi^0_2$-space,
hence a $\Gamma$-space,
we can use Theorem~\ref{th:generalized:lav}
(applied to $X'=Y'=\calN$)
to extend the homeomorphism $s:Y\to S$ to a homeomorphism $s^\ast$
between larger sets $Y^\ast,S^\ast\in\bfPi^0_2(\calN)$.
By Proposition~\ref{propbh} we have
$Y^\ast,S^\ast\in\bfPi^0_\alpha(\calN)$.
Since $S\in\bfPi^0_\alpha(S^\ast)$ by Proposition \ref{sub},
$Y=s^{\ast-1}(S)\in\bfPi^0_\alpha(Y^\ast)$.
Therefore $Y\in\bfPi^0_\alpha(\calN)$ by Proposition~\ref{sub},
a contradiction.
$\Box$

\medskip

\begin{remark}
 The spaces $Y$ witnessing the non-collapse
 property above are rather artificial.
 In Theorem~\ref{th:luzrep}
 we will find very natural spaces witnessing the non-collapse of
 the Luzin hierarchy of $\QCBZ$-spaces.
\end{remark}

%
%

\section{Relating the Hierarchies}\label{relat}

In this section we establish close relationships of the
hierarchies of $\CBZ$-spaces with the corresponding hierarchies of
$\QCBZ$-spaces. The next result implies that any level of a
hierarchy of $\QCBZ$-spaces extends the corresponding level in the
corresponding hierarchy of $\CBZ$-spaces.

\begin{proposition}\label{extend}
 Let
 $\Gamma \in
  \{\bfPi^0_2,\bfSig^0_\alpha, \bfPi^0_\alpha,\bfSig^1_n, \bfPi^1_n
    \mid 3\leq\alpha<\omega_1,1\leq n<\omega\}$.
  Then $\CBZ(\Gamma)\subseteq \QCBZ(\Gamma)$.
\end{proposition}

{\bf Proof.} It suffices to show that any space
$X\in\Gamma(\Pomega)$ is a $\Gamma$-representable space. Let
$\rho:\calN\to \Pomega$ be the total admissible representation of
$\Pomega$ defined by $\rho(x)=\{n\mid \exists i(x(i)=n+1)\}$ (see
e.g. \cite{wei00,br} for details). Then the restriction of $\rho$
to $\rho^{-1}(X)\in\Gamma(\calN)$ is an admissible representation
of $X$. Since $X$ is a $\CBZ$-space, $X$ is a
$\Gamma$-representable space by
Proposition~\ref{p:equiv:Gamma-representable}. $\Box$

In particular, we have $\CBZ(\Gamma)\subseteq \QCBZ(\Gamma)\cap\CBZ$
for each level $\Gamma$ of the Borel hierarchy or the Luzin hierarchy.
The main question of this section is: for which levels $\Gamma$ we
have the equality $\CBZ(\Gamma)= \QCBZ(\Gamma)\cap \CBZ$?
Proposition \ref{proppi2} implies that the equality holds for
$\Gamma=\bfPi^0_2$.

The next result implies that the equality holds for all
zero-dimensional $\CBZ$-spaces.

\begin{proposition}\label{relzero}
 Let $\Gamma\in\{\bfPi^0_2, \bfSig^0_\alpha, \bfPi^0_\alpha,
 \bfSig^1_n, \bfPi^1_n\mid 3\leq\alpha<\omega_1,1\leq n<\omega\}$
 and $X$ be a zero-dimensional space in $\QCBZ(\Gamma)\cap \CBZ$.
 Then $X\in \CBZ(\Gamma)$.
\end{proposition}

{\bf Proof.} Let $r:D\to X$ be an admissible representation of $X$
with $D\in\Gamma(\calN)$. By Lemma~\ref{l:retr2} there is a
continuous function $s:X\to D$ satisfying $rs=id_X$. Then
$s(X)=\{z\in D\mid sr(z)=z\}\in\bfPi^0_2(D)$ by
Proposition~\ref{equal}, hence $s(X) \in \Gamma(\calN)$ by
Proposition~\ref{sub}. Since $\calN$ is a $\bfPi^0_2$-space, there
is a homeomorphism $f$ of $\calN$ onto a subspace $Y$ of $\Pomega$
with $Y \in \bfPi^0_2(\Pomega)$. As $fs$ is a homeomorphism of $X$
onto $fs(X)$, we have $fs(X)\in\Gamma(Y)$.
Propositions~\ref{propbh} and~\ref{sub} yield
$fs(X)\in\Gamma(\Pomega)$. Therefore $X\in \CBZ(\Gamma)$. $\Box$

By the next theorem the equality $\CBZ(\Gamma)= \QCBZ(\Gamma)\cap
\CBZ$ holds for almost all levels. For finite levels, this was
pointed out to us by Matthew de Brecht. We thank him for giving
the permission to use his proof of Theorem
\ref{th:relat:levels}(2). Note that his proof may be used to
obtain also a proof of Theorem \ref{th:relat:levels}(1) which is
slightly different from ours.

\pagebreak[3]
\begin{theorem} \label{th:relat:levels}
\begin{description}
\item[\rm(1)\,]
 For any level
 $\Gamma \in \{\bfSig^0_\alpha, \bfPi^0_\alpha, \bfSig^1_n, \bfPi^1_n \mid
   \omega\leq\alpha<\omega_1, 1\leq n<\omega\}$, we have
 \[
   \QCBZ(\Gamma) \cap \CBZ = \CBZ(\Gamma) \,.
 \]
\item[\rm(2)\,]
 For all natural numbers $m \geq 2$ and $n \geq 3$, we have
 \[
  \QCBZ(\bfPi^0_m)\cap \CBZ = \CBZ(\bfPi^0_{m})
  \quad\text{and}\quad
  \QCBZ(\bfSig^0_n)\cap \CBZ = \CBZ(\bfSig^0_{n}) \,.
 \]
\end{description}
\end{theorem}

\pagebreak[3] {\bf Proof.}
\begin{description}
\item[\rm(1)\,]
 Let $\Gamma\in\{\bfSig^0_\alpha, \bfPi^0_\alpha, \bfSig^1_n, \bfPi^1_n
  \mid \omega \leq\alpha<\omega_1,1\leq n<\omega\}$.
 We have already seen $\CBZ(\Gamma) \subseteq \QCBZ(\Gamma) \cap \CBZ$.

 Let $X\in \QCBZ(\Gamma)\cap \CBZ$. Then there is an
 admissible representation $\delta:D\to X$ of $X$ with
 $D\in\Gamma(\calN)$. By Proposition \ref{p:inj1} we may
 assume w.l.o.g. $X$ to be a subspace of $\Pomega$.
 We have to show $X \in \CBZ(\Gamma)$.

 Let $\chi:\Pomega\to\mathcal{C}$ be the bijection between
 $\Pomega$ and the Cantor space $\mathcal{C}=2^\omega$ that sends
 any $A\subseteq\omega$ to its characteristic function $\chi_A$.
 Obviously, $\chi$ is $\bfSig^0_2$-measurable while its
 inverse $\chi^{-1}$ is continuous (i.e.
 $\bfSig^0_1$-measurable). Let $\sigma=\chi|_X$ be the
 restriction of $\chi$ to $X\subseteq \Pomega$. By Proposition
 \ref{sub}, $\sigma$ is a $\bfSig^0_2$-measurable
 bijection between $X$ and $\sigma(X)\subseteq \mathcal{C}$ such
 that $\sigma^{-1}$ is $\bfSig^0_1$-measurable.

 Since $\sigma^{-1}$ is continuous as a function
 from the zero-dimensional $\CBZ$-space $\sigma(X)$ to $X$
 and $\delta$ is admissible, there is a continuous function
 $g\colon \sigma(X) \to \calN$ with $\sigma^{-1}=\delta g$.
 Then $id_X=\sigma^{-1}\sigma=\delta g\sigma$, hence the function
 $h=g\sigma$ satisfies $h(X)=\{y\in D\mid h\delta(y)=y\}$. Thus,
 $h$ is a $\bfSig^0_2$-measurable bijection between $X$
 and $H=h(X)$, whereas its inverse $h^{-1}=\delta|_{h(X)}$ is
 $\bfSig^0_1$-measurable (i.e.\ continuous).

 By Proposition \ref{relzero} we may w.l.o.g.\ assume that $D$ is a
 subspace of $\Pomega$ and $D\in\Gamma(\Pomega)$. Since $H$ is the
 preimage of $EQ_D\in\bfPi^0_2(D\times D)$ under the
 $\bfSig^0_2$-measurable function $y\mapsto (h\delta(y),y)$,
 we have $H\in\bfPi^0_3(D)$ by Proposition~\ref{meas}
 and hence $H \in \Gamma(D)$ by Proposition~\ref{propbh}.
 Proposition~\ref{sub} yields $H\in\Gamma(\Pomega)$.

 By Theorem \ref{lavrPo} and Corollary \ref{lavrPo1}, there exist
 $H^\ast\in\bfPi^0_3(\Pomega)$,
 $X^\ast\in\bfPi^0_4(\Pomega)$, a continuous bijection
 $\delta^\ast:H^\ast\to X^\ast$ and a
 $\bfSig^0_2$-measurable bijection $h^\ast:X^\ast\to
 H^\ast$ such that $\delta^\ast$ is the inverse of $h^\ast$,
 $\delta^\ast$ is an extension of $h^{-1}$ and $h^\ast$ is an
 extension of $h$.
 Since $H\in\Gamma(H^\ast)$ by Proposition \ref{sub} and $X=h^{\ast-1}(H)$,
 Proposition~\ref{meas} yields us $X\in\Gamma(X^\ast)$.
 By Propositions~\ref{propbh} and~\ref{sub}, we obtain
 $X\in\Gamma(\Pomega)$, hence $X\in \CBZ(\Gamma)$.
 This completes the proof.
\item[\rm(2)\,]
 Let $\delta$ be an admissible representation
 of a subspace $X$ of $\Pomega$ such that
 $\dom(\delta) \in \Gamma(\calN)$, where
 $\Gamma \in \{ \bfPi^0_m, \bfSig^0_n\,|\, m \geq 2,\, n \geq 3 \}$.
 Let $\rho$ be the total admissible representation of $\Pomega$
 defined in the proof of Proposition~\ref{extend}.
 By Proposition~\ref{p:BrYa} it is enough
 to show that $\rho^{-1}(X)$ is in $\Gamma(\calN)$.
 Since the corestriction of $\rho$ to $X$ is continuous,
 there is a continuous function $h\colon \rho^{-1}(X) \to \calN$
 satisfying $\rho(p)=\delta h(p)$ for all $p \in \rho^{-1}(X)$.
 As $\Pomega$ is an injective space and $\calN$ is a Polish space,
 $\delta$ and $h$ can be extended to continuous functions
 $\delta^*\colon \calN \to \Pomega$ and $h^* \colon G \to \calN$,
 where $G$ is a $\bfPi_2^0$-subset of $\calN$,
 see Propositions~\ref{inject} and~\ref{p:ext:completely}.
 By Proposition~\ref{equal},
 $A:=\{ p \in G \,|\, \rho(p)=\delta^*h^*(p)\}$ is
 a $\bfPi_2^0$-subset of $G$ and thus of $\calN$.
 It is easy to verify that $\rho^{-1}(X)$
 is the intersection of $A$ with $(h^*)^{-1}(\dom(\delta))$.
 By Proposition~\ref{clos} and~\ref{sub} this implies
 that $\rho^{-1}(X)$ is $\Gamma$-subset of $\calN$.
 Proposition~\ref{p:BrYa} yields that
 $X$ is a $\Gamma$-subset of $\Pomega$, i.e., $X \in \CBZ(\Gamma)$.
\end{description}
$\Box$

%
%

\section{The Luzin Hierarchy and Continuous Functionals}\label{functional}

In this section we establish close relations of the Luzin hierarchy
to the continuous functionals of finite types.

We start by relating the exponentiation operation on admissibly
represented spaces (see Proposition \ref{fspace}) to the Luzin hierarchy.

\begin{theorem}\label{th:exp}
 Let $k \in \omega$,
 $X\in \QCBZ(\bfPi^1_k)$ and $Y\in \QCBZ(\bfSig^1_k)$.
 Then $Y^X\in \QCBZ(\bfPi^1_{k+1})$.
\end{theorem}

{\bf Proof.} Let $\delta$ and $\gamma$ be admissible
representations of $X$ and $Y$ respectively such that
$\mathit{EQ}(\delta)\in\bfPi^1_k(\calN^2)$ and
$\mathit{EQ}(\gamma)\in\bfSig^1_k(\calN^2)$. By Proposition
\ref{fspace} it suffices to show that
$\mathit{EQ}([\delta\to\gamma])\in\bfPi^1_{k+1}(\calN^2)$.

From the definition of $[\delta\to\gamma]$ we obtain
\[
  (p,q)\in \mathit{EQ}([\delta\to\gamma])
  \Longleftrightarrow \left\{
  \begin{array}{l}
       \text{for all $(x,y) \in \mathit{EQ}(\delta)$}
     \\
      \big(u_p(x),u_p(y)\big),
      \big(u_{q}(x),u_{q}(y)\big),
      \big(u_{p}(x),u_{q}(x)\big)
      \in \mathit{EQ}(\gamma).
  \end{array}\right.
\]
So we have
\[
 \mathit{EQ}([\delta\to\gamma]) =
 \big\{ (p,q) \in \calN^2 \,\big|\,
   \forall (x,y) \in \calN^2. (p,q,x,y) \in M
 \big\} \,,
\]
where
\[
 M:= \Big\{ \begin{array}[t]{@{\,}l}
     (p,q,x,y) \in \calN^4\,\Big|\,
     (x,y) \notin \mathit{EQ}(\delta)
     \text{ or there is some $(a,b,c,d) \in \calN^4$ such that}
    \\
      (p,x,a), (p,y,b), (q,x,c), (q,y,d) \in \graph(u)
       \text{ and } (a,b), (c,d), (a,c) \in \mathit{EQ}(\gamma)
    \Big\}.
   \end{array}
\]
Since the universal function $u :\subseteq \calN^2 \to \calN$
is continuous and has a $\bfPi^0_2$-set as domain,
$\graph(u)$ is a $\bfPi^0_2(\calN^3)$-set.
This implies that $M$ is a $\bfSig^1_n(\calN^4)$-set
by Propositions~\ref{propbh} and~\ref{clos}.
We conclude that $\mathit{EQ}([\delta\to\gamma])$
is a $\bfPi^1_{n+1}(\calN^2)$-set.
$\Box$

The next result (which is known from \cite{kr59} for the
particular case $Y=\Nk{k}$) is the key technical fact of this
section.


\begin{theorem}\label{th:key}
 Let $Y$ be a $\QCBZ$-space and let $f\colon Y\to\calN$ be
 a continuous function with $\rng(f) \neq \calN$.
 Then there exists
 a continuous function $g\colon \calN \times \omega^Y\to\calN$
 with $\rng(g)=\calN\setminus \rng(f)$.
\end{theorem}

We remark that by $\calN \times \omega^Y$ we mean the product
formed in the category of $\QCBZ$-spaces (see Subsection~\ref{admiss}),
the topology of which is finer than (or equal to) the Tychonoff topology.
However, the function $g$ constructed in the proof is even
continuous w.r.t.\ the Tychonoff topology.

{\bf Proof.}
 We abbreviate $M=\calN\setminus \rng(f)$.
 The set
 $\mathit{NEQ}=
  \big\{ (x,y) \in \calN \times Y \,\big|\, f(y) \neq x \big\}$
 is the countable union of the clopen sets
 \[
  \{ x \in \calN \,|\, x(j)=a \}
  \times
  f^{-1}\{ z \in \calN \,|\, z(j)=b\} \,,
 \]
 where $(j,a,b)$ varies over all triples of natural numbers
 with $a \neq b$.
 Let $\{D_i\}_i$ denote a sequence consisting of these clopen sets.
 So we have $\mathit{NEQ}=\bigcup_{i \in \omega} D_i$
 and
 $x \in M \Longleftrightarrow
  \forall y \in Y. \exists i \in \omega. (x,y) \in D_i$.

 Motivated by the above equivalence,
 we call a continuous function $H\colon Y \to \omega$
  \emph{a witness for} an element $x \in M$, if $(x,y) \in D_{H(y)}$
 holds for all $y \in Y$.
 The idea of the proof is to construct the required function
 $g\colon \calN \times \omega^Y \to \calN$ in such a way
 that $g$ maps $(x,H)$ to $x$, if $H$ is a witness for $x$.
 Otherwise $g$ assigns to $(x,H)$ some element
 of a countable dense subset of $M$.

 To construct an appropriate dense subset of $M$,
 we choose for every finite sequence $w$
 in the set $W:=\{ x^{<k} \,|\, x \in M ,\, k \in \omega \}$
 some $\alpha(w) \in M$ such that $\alpha(w)$
 has $w$ as an initial segment.
 Clearly, $\{ \alpha(w) \,|\, w \in W\}$ is dense in~$M$.

 By being a $\QCBZ$-space,
 $Y$ has a countable dense subset $\{ \beta_j \,|\, j \in \omega\}$
 (see Proposition 3.3.1 in~\cite{sch03}).
 We construct a sequence $\{C_k\}_k$ of subsets
 of the $\QCBZ$-space $\calN \times \omega^Y$ by
 $C_{0}:=\calN \times \omega^Y$ and
 \[
  C_k:=\big\{ (x,H) \in \calN \times \omega^Y
     \,\big|\, x^{<k} \in W \text{ and }
    (x,\beta_j) \in D_{H(\beta_j)} \text{ for all $j \leq k$} \big\}
 \]
 for all $k > 0$.
 The set $C_k$ is clopen in the Tychonoff topology on
 $\calN \times \omega^Y$ and thus in the $\QCBZ$-topology,
 because $C_k$ is the union of the clopen sets
 \[
  \big( (w\cdot\calN) \times \omega^Y \big)
  \cap
  \bigcap_{j=0}^k
    \big( \{x \in \calN \,|\, (x,\beta_j) \in D_{a_j} \}
     \times
     \{ H \in \omega^Y \,|\, \beta_j \in H^{-1}\{a_j\} \} \big),
 \]
 where $w$ ranges over the finite sequence of length $k$ in $W$
 and $a_0,\dotsc,a_k$ vary over the natural numbers.
 Note that $\{ H \in \omega^Y \,|\, \beta_j \in H^{-1}\{a_j\} \}$
 is clopen even in the compact-open topology on $\omega^Y$ and
 therefore in the $\QCBZ$-topology on $\omega^Y$, which is finer
 than the former.

 Next we define a function
 $\ell\colon \calN \times \omega^Y \to \omega \cup \{\infty\}$
 by
 \[
   \ell(x,H):= \left\{
   \begin{array}{cl}
     \infty & \text{if $(x,H) \in \bigcap_{k \in \omega} C_k$}
     \\
     \max\{ k \in \omega \,|\, (x,H) \in C_k \} & \text{otherwise}
   \end{array} \right.
 \]
 We use $\{C_k\}_k$ and $\ell$ to define
 our function $g: \calN \times \omega^Y \to \calN$ by
 \[
   g(x,H):= \left\{
   \begin{array}{ll}
     x & \text{if $(x,H) \in \bigcap_{k \in \omega} C_k$}
     \\
     \alpha(x^{<\ell(x,H)}) & \text{otherwise}
   \end{array}\right.
 \]
 Note that $\infty \neq i \leq \ell(x,H)$ implies $x^{<i} \in W$
 and $g(x,H)^{<i}=x^{<i}$.

 First we show the continuity of $g$.
 Let $O$ be an open set of the Baire space
 and let $(x,H) \in g^{-1}(O)$.
 Then there is some $m \in \omega$ such that
 $g(x,H)^{<m}\cdot \calN \subseteq O$.
 \smallskip\\
 Case 1:
  Assume $\ell(x,H)=\infty$. Then $(x,H) \in \bigcap_{k\in\omega} C_k$.
  We define a set $U$ by
  \[
   U:= (x^{<m}\cdot\calN \times \omega^Y)
     \cap \bigcap_{i=0}^{m} C_i
  \]
  Then $U$ is open and all $(x',H') \in U$ satisfy
  $\ell(x',H')\geq m$ and $g(x',H')^{<m}=g(x,H)^{<m}=x^{<m}$.
  Hence $(x,H) \in U \subseteq g^{-1}(O)$.
 \smallskip\\
 Case 2: Assume $\ell(x,H) < \infty$.
  Let $k:=\ell(x,H)$ and define a set $U$ by
  \[
   U:= (x^{<k}\cdot\calN \times \omega^Y) \cap
     \bigcap_{i=0}^{k} C_i \setminus C_{k+1}
   \]
  Then $U$ is open, $(x,H) \in U$ and all $(x',H') \in U$
  satisfy even $g(x',H')=g(x,H)=\alpha(x^{<k}) \in U$.
 \smallskip\\
 We conclude that in both cases $g$ is continuous in the point $(x,H)$.
 Therefore $g$ is continuous, even with respect to the Tychonoff product
 on the set $\calN \times \omega^Y$, which is coarser than
 the $\QCBZ$-topology.

 Now we show $M \subseteq \rng(g)$.
 Let $x \in M$. We define a function $H_x\colon Y \to \omega$ by
 \[
   H_x(y):=\min\{ k \in \omega \,|\, (x,y) \in D_k\} \,.
 \]
 Note that $H_x$ is total, because we have $f(y) \neq x$ and
 therefore $(x,y) \in \bigcup_{k \in \omega} D_k$ for every $y \in Y$.
 Furthermore, $H$ is continuous in every point $y \in Y$,
 because the set
 \[
  U:=
   \big\{ b \in Y \,\big|\,
       (x,b) \in D_{H_x(y)} \setminus
       \mathop{\textstyle\bigcup}_{i=0}^{H_x(y)-1} D_i \big\}
 \]
 satisfies $y \in U \subseteq H_x^{-1}\{ H_x(y)\}$
 and is open by being the preimage of the open set
 $D_{H_x(y)} \setminus \bigcup_{i=0}^{H_x(y)-1} D_i$
 under the continuous map $b\mapsto (x,b)$.
 Therefore $H_x$ is an element of the function space $\omega^Y$.
 Since $(x,H_x) \in \bigcap_{k \in \omega} C_k$, we have $g(x,H_x)=x$.

 It remains to show $\rng(g) \subseteq M$.
 So let $x \in \rng(g)$.
 The only interesting case is $x \notin \{ \alpha(w) \,|\, w \in W\}$.
 Then there is some continuous function $H\colon Y \to \omega$
 such that $g(x,H)=x$ and thus $(x,H) \in \bigcap_{k \in \omega} C_k$.
 Suppose for a contradiction that there is some $y \in Y$ such that
 $(x,y) \notin D_{H(y)}$.
 Then the set
 \[
  V:=\{ b \in Y \,|\, (x,b) \notin D_{H(y)}\}
     \cap H^{-1}\{ H(y) \}
 \]
 is non-empty and open in $Y$, because $D_{H(y)}$ is closed and $H$
 is continuous.
 So there exists some $k\in \omega$ with $\beta_k \in V$.
 The element $\beta_k$ satisfies
 $(x,\beta_k) \notin D_{H(y)}$ and $H(\beta_k)=H(y)$.
 But since $(x,H) \in C_k$, we have $(x,\beta_k) \in D_{H(\beta_k)}$,
 a contradiction.
 \\
 We conclude $(x,y) \in D_{H(y)}$ and thus $f(y) \neq x$
 for every $y \in Y$.
 Therefore $x \notin \rng(f)$ and $x \in M$.

 So $g$ satisfies the required properties.
$\Box$

Using the cartesian closedness of $\QCBZ$, we define a sequence of
spaces $\{\Nk{k}\}_{k<\omega}$ by induction on $k$ as follows:
$\Nk{0}:=\omega$ and $\Nk{k+1}:=\omega^{\Nk{k}}$, where $\omega$
denotes the space of natural numbers endowed with the discrete
topology. Obviously $\Nk{1}$ is equal to the Baire space $\calN$.
The space $\Nk{k}$ is referred to as \emph{the sequential space of
(Kleene-Kreisel) continuous functionals of type $k$}.
For any $k \geq 2$, the sequential topology on $\Nk{k}$ is strictly finer
than the corresponding compact-open topology \cite{hyland}.
Furthermore it is neither zero-dimensional nor regular \cite{Sch:NNN}.
Any of these spaces has a natural admissible representation.
From Theorem~\ref{th:exp} we obtain:

\begin{corollary}\label{c:levels:of:Nk(k)}
 For any integer $k \geq 1$,
 $\Nk{k} \in \QCBZ(\bfPi^1_{k-1})$.
\end{corollary}

{\bf Proof.} Obviously,
 $\mathbb{N}\langle0\rangle,\mathbb{N}\langle 1\rangle \in
  \QCBZ(\bfPi^0_0) \subseteq \QCBZ(\mathbf{\Delta}^1_1)$.
 By Theorem \ref{th:exp}, $\Nk{2}\in \QCBZ(\bfPi^1_1)$.
 Assuming by induction that $\Nk{k}\in \QCBZ(\bfPi^1_{k-1})$
 for a given $k\geq2$, we obtain $\Nk{k+1}\in \QCBZ(\bfPi^1_k)$ by
 Theorem~\ref{th:exp}.
$\Box$

We remark that Lemma 2.34 in \cite{no80}
implies that $\Nk{k}$ has an admissible representation
such that its domain is even in the effective version $\Pi^1_{k-1}$
of $\bfPi^1_{k-1}(\calN)$.
This follows from the fact that the function which maps
``associates'' for $\Nk{k}$ to the encoded functionals
forms an admissible representation.

Note that the continuous functionals
(also known under the somewhat misleading name ``countable functionals'')
were first defined in \cite{kl59,kr59} independently
by S. Kleene and G. Kreisel. Their definitions look different from
each other, as well as from the definition above, although all
three definitions are known to be equivalent. Another equivalent
definition in terms of domains was given in \cite{er74}.
Additional information may be found in \cite{no80,no99}.

The following observation follows from the cartesian closedness
of $\QCBZ$.

\begin{lemma}\label{l:N<k>retract}
 For $k \geq 1$, the $\QCBZ$-spaces $\Nk{k-1}$ and $\calN \times \Nk{k}$
 are continuous retracts of $\Nk{k}$.
\end{lemma}

{\bf Proof.}
 As $\QCBZ$ is a cartesian closed category,
 for all $\QCBZ$-spaces $X,Y,Z$ such that $X$ is a continuous retract
 of $Y$ the function space $Z^X$ is a continuous retract of $Z^Y$
 and the $\QCBZ$-product $X \times Z$ is a continuous retract of
 $Y \times Z$.
 Clearly, $\omega$ is a continuous retract of $\calN=\Nk{1}$.
 So the first statement follows by induction on $k$.
 A further induction establishes $\calN$ as a continuous retract
 of $\Nk{k}$ for $k \geq 1$.
 Hence $\calN \times \Nk{k}$ is a continuous retract
 of $\Nk{k} \times \Nk{k}$.
 The cartesian closedness of $\QCBZ$ allows us to calculate
 $$\Nk{k}
  = \omega^{\Nk{k-1}}
  \cong (\omega \times \omega)^{\Nk{k-1}}
  \cong \omega^{\Nk{k-1}} \times \omega^{\Nk{k-1}}
  = \Nk{k} \times \Nk{k},
 $$
 hence $\Nk{k}$ and $\Nk{k} \times \Nk{k}$ are isomorphic in $\QCBZ$
 and thus homeomorphic.
 We conclude that $\calN \times \Nk{k}$ is a continuous retract
 of $\Nk{k}$.
$\Box$

From Theorem~\ref{th:key} and Lemma~\ref{l:N<k>retract}
we can easily infer the following nice result.

\begin{proposition}\label{p:rng}
 For any positive integer $k$ and for any non-empty set
 $M \in\bfSig^1_k(\calN)$ there is a continuous
 function $f:\Nk{k}\to \calN$ with $\rng(f)=M$.
\end{proposition}

{\bf Proof.}
 We proceed by induction on $k \geq 1$.
 For $k=1$ the claim is well-known (cf.\ Section~14 in~\cite{ke95}).
\\
 Let $M \in \bfSig^1_{k+1}(\calN)$.
 Then there is some set $A \in \bfPi^1_k(\calN)$
 such that
 \[
  M=\big\{ x \in \calN \,\big|\,
    \exists p \in \calN. \langle p,x \rangle \in A \big\} \,,
 \]
 where $\langle \cdot,\cdot \rangle$ denotes a canonical homeomorphism
 from $\calN^2$ to $\calN$.
 We set $B:=\calN \setminus A$, hence $B \in \bfSig^1_k(\calN)$.
 Since $M \in \{ \emptyset, \calN \}$ if $A\in \{ \emptyset, \calN \}$,
 we can assume $\emptyset \neq A,B \neq \calN$.
 The induction hypothesis yields us a continuous function
 $f_B\colon \mathbb{N}\langle k\rangle \to \calN$ such that $\rng(f_B)=B$.
 By Theorem~\ref{th:key} there exists a continuous function
 $g\colon \calN \times \Nk{k+1} \to \calN$
 such that $\rng(g)=A$.
 By Lemma~\ref{l:N<k>retract} there is a continuous retraction
 $r\colon \Nk{k+1} \to \calN \times \Nk{k+1}$.
 Since $r$ is surjective, $A$ is also the range of $gr$.
 Using the unique continuous function $\pi_2^{(2)}\colon \calN \to \calN$
 satisfying $\pi_2^{(2)}\langle p,x \rangle=x$,
 we define $f\colon \Nk{k+1} \to \calN$ by
 $f(z):=\pi_2^{(2)}(gr(z))$.
 Then $f$ is continuous and satisfies
 \begin{align*}
  x \in M
  &\iff
  \exists p \in \calN.\, \langle p,x \rangle \in A
  \iff
  \exists \phi \in \Nk{k+1}. \exists p \in \calN.\,
    gr(\phi)=\langle p,x \rangle
  \\
  &\iff
  \exists \phi \in \Nk{k+1}.\, f(\phi)=x \,.
 \end{align*}
 Hence $M$ is the range of $f$.
$\Box$

The last result can be even improved to the following
characterization of levels of the Luzin hierarchy in terms of the
Kleene-Kreisel continuous functionals.

\begin{theorem}\label{charluz}
 Let $k$ be a positive integer and $B$ a non-empty subset of $\calN$.
 Then $B\in\bfSig^1_k(\calN)$ iff
 there is a continuous function $f\colon \Nk{k}\to \calN$
 with $\rng(f)=B$.
\end{theorem}

{\bf Proof.}
 The only-if-part is given by Proposition~\ref{p:rng}.
 For the if-part let $f\colon \Nk{k}\to \calN$ be a continuous
 function.
 According to Corollary~\ref{c:levels:of:Nk(k)},
 $\Nk{k}$ has an admissible representation $\delta$
 such that its domain $\dom(\delta)$ is a $\bfPi^1_{k-1}$-set.
 We define a partial function $g\colon \calN \to \calN$ by $g(p):=f\delta(p)$.
 Then $g$ is continuous.
 So it can be extended to a continuous function
 $g^*\colon X \to \calN$ such that $X$ is a $\bfPi^0_2$-subset of $\calN$
 (cf.\ Proposition~\ref{p:univ} or~\ref{p:ext}).
 The graph of $g^*$ is a closed set in $X \times \calN$
 and therefore a $\bfPi^0_2$-set in $\calN \times \calN$.
 By Propositions~\ref{propbh} and~\ref{clos},
 the set $A:=\graph(g^*) \cap (\dom(\delta) \times \calN)$
 is a $\bfPi^1_{k-1}$-subset of $\calN \times \calN$.
 Moreover, we have
 \[
   M=\{ y \in \calN \,|\, \exists x \in \calN.\, (x,y) \in A \}
    = \mathit{pr}_\calN(A)
   \,.
 \]
 Therefore $M \in \bfSig^1_k(\calN)$.
$\Box$

A similar characterization for the effective version of
$\bfSig^1_k(\calN)$ can be found in \cite[Theorem 5.22]{no80}.

Finally, we relate continuous functionals to the Luzin hierarchy of
$\QCBZ$-spaces (for a similar relationship see \cite{no81}). The
next result provides the exact estimation of the spaces of
continuous functionals of finite types in the Luzin hierarchy of
$\QCBZ$-spaces. On the other hand, the result provides ``natural''
witnesses for the non-collapse property of this hierarchy.

\begin{theorem}\label{th:luzrep}
 For any positive integer $k$,
 $\Nk{k+1} \in \QCBZ(\bfPi^1_k)\setminus \QCBZ(\bfSig^1_k)$.
\end{theorem}

{\bf Proof.} We know $\Nk{k+1} \in \QCBZ(\bfPi^1_k)$ from
Corollary~\ref{c:levels:of:Nk(k)}. It remains to show that
$\mathbb{N}\langle k+1\rangle\not\in \QCBZ(\bfSig^1_k)$. We prove
the stronger assertion that $\Nk{k+1}$ has no continuous
representation $\delta$ with $\dom(\delta)\in\bfSig^1_k(\calN)$.

Suppose for a contradiction that $\delta$ is a continuous representation
of $\Nk{k+1}$ with $\dom(\delta) \in \bfSig^1_k(\calN)$.
Then there is a continuous function $f\colon \Nk{k} \to \calN$ with
$\rng(f)=\dom(\delta)$ by Proposition~\ref{p:rng}.
We define a function $g\colon \Nk{k} \to \omega$ by
\[
 g(\phi)
 := 1 + \delta(f(\phi))(\phi)
  = 1 + \mathit{eval}\big( \delta(f(\phi)),\phi \big)\,.
\]
Then $g$ is continuous, because $f$, $\delta$ and the evaluation function
$\mathit{eval}\colon \Nk{k+1} \times \Nk{k} \to \omega$
mapping $(\psi,\phi)$ to $\psi(\phi)$ are continuous
(note that $\Nk{k+1} \times \Nk{k}$ carries
the sequential $\QCBZ$-topology).
So there is some $p \in \dom(\delta)$ and some $h \in \Nk{k}$
such that $\delta(p)=g$ and $f(h)=p$.
We obtain
\[
 g(h)= 1 + \delta(f(h))(h) = 1 + g(h)\,,
\]
a contradiction.
\\
We conclude $\Nk{k+1} \notin \QCBZ(\bfSig^1_k)$.
$\Box$

%

\section{The category of projective $\QCBZ$-spaces}
\label{categ}

In this section we show that the Luzin hierarchy of $\QCBZ$-spaces
gives rise to a nice cartesian closed category. This is the full
subcategory of the category $\QCBZ$ consisting of the
$\bigcup_n\QCBZ(\bfSig^1_n)$ as objects and all continuous
function between them as morphisms. We denote this category by
$\QCBZ(\mathbf{P})$ and call its objects \emph{projective
qcb-spaces}.

\begin{theorem}\label{th:QCBZ(P):is:ccc}
 The category $\QCBZ(\mathbf{P})$ of projective qcb-spaces is cartesian closed.
\end{theorem}

{\bf Proof.} We only have to check that $\QCBZ(\mathbf{P})$ is
closed under binary products and exponentiation formed in the
supercategory $\QCBZ$. For exponentiation this follows from
Theorem \ref{th:exp}. It is an easy exercise to show that the
product representation $[\delta \times \gamma]$ constructed in
Proposition~\ref{fspace} satisfies $\mathit{EQ}([\delta \times
\gamma]) \in \bfSig^1_n(\calN^2)$, whenever $\mathit{EQ}(\delta)$
and $\mathit{EQ}(\gamma)$ are $\bfSig^1_n$-subsets of $\calN^2$.
$\Box$

It turns out that $\QCBZ(\mathbf{P})$ is in a sense
the smallest cartesian closed subcategory of $\QCBZ$ containing $\omega$.

\begin{proposition}\label{p:QCBZ(P):is:ccc}
 There is no full cartesian closed subcategory $\sfC$ of $\QCBZ$
 such that $\sfC$ inherits binary products from $\QCBZ$,
 contains the discrete space $\omega$ of natural numbers
 and is contained itself in $\QCBZ(\bfSig^1_n)$
 for some $1 \leq n < \omega$.
\end{proposition}

{\bf Proof.} We show at first that any cartesian closed
subcategory $\sfD$ of $\QCBZ$ that contains $\omega$ and inherits
binary products from $\QCBZ$ has the property that exponentials
formed in $\sfD$ are homeomorphic to the corresponding
$\QCBZ$-exponentials.

Let $E$ be an exponential formed in $\sfD$ to spaces $X,Y \in \sfD$.
Remember that this means that there exists a continuous evaluation function
$\mathit{eval} \colon E \times X \to Y$
such that for every space $Z \in \sfD$
and every continuous function $h\colon Z \times X \to Y$
there is a unique continuous function $\hat{h}\colon Z \to E$
such that
\[
  h(z,x)=\mathit{eval}\big( \hat{h}(z)(x) \big)
  \quad\text{for all $z \in Z$ and $x \in X$.}
\]
 For every continuous function $f\colon X \to Y$
 it follows from the uniqueness condition
 (applied to $Z=\omega$ and the continuous function $(i,x) \mapsto f(x)$)
 that there exists a unique element $t(f) \in E$ satisfying
 $f(x)=\mathit{eval}(t(f),x)$ for all $x \in X$.
 Conversely, since $Y^X$ is an exponential to $X$ and $Y$
 in the supercategory $\QCBZ$,
 there is a continuous function $\mathit{Ev}\colon E \to Y^X$
 satisfying $\mathit{Ev}(e)(x)=\mathit{eval}(e,x)$ for all $e \in E$
 and $x \in X$.
 Clearly, $\mathit{Ev}(t(f))=f$ for every $f \in Y^X$.
 The uniqueness condition implies $t(\mathit{Ev}(e))=e$
 for every $e \in E$. Hence $\mathit{Ev}$ is a continuous bijection
 with $t$ as its inverse.

 To show that $t$ is a continuous function from $Y^X$ to $E$,
 let $\{f_i\}_i$ be a sequence
 that converges in $Y^X$ to some function $f_\infty \in Y^X$.
 This can be reformulated by stating that the function
 $F\colon \INu \times X \to Y$ defined by $F(i,x):=f_i(x)$
 for $i \in \omega \cup \{\infty\}$ and $x \in X$
 is continuous, where $\INu$ denotes the one-point compactification
 of the discrete natural number with $\infty$ as the infinity point.
 (This is a well-known property of exponentials in the category
 of sequential topological spaces and therefore in the category $\QCBZ$,
 see e.g.\ \cite{ELS04,sch03}).

 Below we will show that $\INu$ is a continuous retract
 of the exponential $E_\calN$ formed in $\sfD$ to $X=Y=\omega$.
 So let $(s,r)$ be a continuous section-retraction pair.
 We define $G\colon E_\calN \times X \to \omega$ by $G(p,x):=F(r(p),x)$.
 Clearly, $G$ is continuous.
 Since $E$ is an exponential in $\sfD$,
 there is a continuous function $\widehat{G}\colon E_\calN \to E$
 satisfying $\mathit{eval}(\widehat{G}(p),x)=G(p,x)$
 for all $p \in E_\calN$ and $x \in X$.
 The uniqueness condition implies $\widehat{G}(p)=t(f_{r(p)})$.
 As $\{i\}_i$ converges to $\infty$ in $\INu$,
 $\{\widehat{G}(s(i))\}_i$ converges to $\widehat{G}(s(\infty))$ in $E$.
 Clearly, we have $\widehat{G}(s(i))=t(f_i)$ for all
 $i \in \omega \cup \{\infty\}$.
 Therefore $\{t(f_i)\}_i$ converges to $t(f_\infty)$ in $E$.

 We conclude that $t$ is sequentially continuous.
 Since $Y^X$ is a sequential topological space,
 $t$ is even continuous in the topological sense.
 This means that $Y^X$ is homeomorphic to $E$.

 Now we show that $\INu$ is indeed a retract of $E_\calN$.
 As in the general case, there is a continuous bijection
 $\mathit{Ev}_\calN\colon E_\calN \to \calN$.
 Hence $E_\calN$ is Hausdorff,
 but not discrete by being an uncountable space
 with a dense countable subset (see Proposition~3.3.1 in \cite{sch03}).
 So there exists an injective sequence $\{z_i\}_i$ converging in $E_\calN$
 to some point
 $z_\infty \in E_\calN \setminus \{ z_i \,|\, i \in \omega \}$.
 Since $\calN$ is zero-dimensional and $\mathit{Ev}_\calN$ is injective,
 there is a sequence $\{C_k\}_k$ of clopen sets in $\calN$
 satisfying
 \[
  \mathit{Ev}_\calN(z_k) \in C_k
   \quad\text{and}\quad
   \{ \mathit{Ev}_\calN(z_i) \,|\, i \leq \infty,\, i \neq k \}
   \subseteq \calN \setminus C_k
 \]
 for every $k \in \omega$. We define functions
 $s\colon \INu \to E_\calN$ and $r\colon E_\calN \to \INu$ by
 \[
  s(i):=z_i \;\;\text{and}\;\;
  r(z):=\min\big\{ \infty, k \in \omega \,\big|\,
        z \in \mathit{Ev}_\calN^{-1}(C_k) \big\}\,.
 \]
 Then both functions are continuous and satisfy $rs=\mathit{id}_{\INu}$.
 Hence $\INu$ is a continuous retract of $E_\calN$.

Finally, suppose that $\sfC$ were a cartesian closed subcategory
of $\QCBZ$ with the desired properties.
By the above statement,
$\sfC$ contains a space homeomorphic to $\Nk{k}$ for all $k \in \omega$.
In particular, there is a space $E \in \sfC$ homeomorphic to $\Nk{n+1}$.
Hence $\Nk{n+1}$ has an admissible representation $\delta$
such that $\mathit{EQ}(\delta) \in \bfSig^1_n(\calN^2)$.
This contradicts Theorem~\ref{th:luzrep}.
$\Box$

\begin{remark} The assumption $\omega\in \sfC$ in the last proposition
 is essential, because several cartesian closed categories
 of $\omega$-algebraic domains are known \cite{ju90},
 which are all $\QCBZ(\bfPi^0_2)$-spaces by being quasi-Polish.
\end{remark}


\section{Conclusion}\label{con}

Hopefully, the results of this paper show that the introduced
hierarchies are natural and interesting for CA and DST. The study
of these hierarchies is of course in the very beginning and many
natural questions remain open. In particular, we would like to see
more natural and important witnesses for the non-collapse property
of the hierarchies. A systematic development of DST for the
introduced classes of spaces (in particular, for the projective
$\QCBZ$-spaces) seems also a natural direction of future research.

\subsection*{Acknowledgements}
  We thank the referees for valuable remarks.
  We are grateful to Matthew de Brecht for permission
  to include his proof of Theorem~\ref{th:relat:levels}(2).

%

\end{document}